\documentclass[journal,twoside,web]{ieeecolor}
\usepackage{generic}
\usepackage{cite}
\usepackage{amsmath,amssymb,amsfonts}
\usepackage{graphicx}
\usepackage{textcomp}
\usepackage{multirow}
\usepackage{algorithm}
\usepackage{algpseudocode}
\usepackage{listings}
\usepackage{color}
\usepackage{subfig}
\usepackage{booktabs}
\usepackage{url}
\usepackage[bookmarks=false]{hyperref}
\hypersetup{colorlinks=false,pdfborder={0 0 0},hypertexnames=false}
\usepackage[binary-units,per-mode=symbol]{siunitx}
\usepackage[acronym,toc,automake=true]{glossaries}

\makeglossaries % Must be after the acronyms, but Overleaf doc says the opposite

\newacronym{soc}{SoC}{System-on-Chip}
\newacronym{hw}{HW}{hardware}
\newacronym{sw}{SW}{software}
\newacronym[plural=WSNs,firstplural=wearable sensor nodes (WSNs)]{wsn}{WSN}{wearable sensor node}
\newacronym[plural=CVDs,firstplural=cardiovascular diseases (CVDs)]{cvd}{CVD}{cardiovascular disease}
\newacronym{ulp}{ULP}{ultra-low power}
\newacronym{dma}{DMA}{direct memory access}
\newacronym[plural=CGRAs,firstplural=coarse-grained reconfigurable arrays (CGRAs)]{cgra}{CGRA}{coarse-grained reconfigurable array}
\newacronym{paf}{PAF}{paroxysmal atrial fibrillation}
\newacronym[plural=NCDs,firstplural=noncommunicable diseases (NCDs)]{ncd}{NCD}{noncommunicable disease}
\newacronym{beatcl}{BeatCl}{the beat classifier}
\newacronym{cl}{CL}{cluster}
\newacronym{fc}{FC}{fabric controller}
\newacronym{icg}{ICG}{impedance cardiogram}
\newacronym{ipc}{IPC}{Instructions-Per-Cycle}
\newacronym{ippg}{iPPG}{imaging photoplethysmography}
\newacronym{isa}{ISA}{Instruction Set Architecture}
\newacronym{mc}{MC}{main core}
\newacronym{mf}{MF}{morphological filtering}
\newacronym{bpf}{BPF}{band-pass filtering}
\newacronym{fir}{FIR}{finite impulse response}
\newacronym[plural=RCs,firstplural=reconfigurable cells (RCs)]{rc}{RC}{reconfigurable cell}
\newacronym{relen}{Rel-En}{Relative-Energy}
\newacronym{rms}{RMS}{root-mean-square}
\newacronym{scm}{SCM}{standard cell memory}
\newacronym[longplural={static random access memories}]{sram}{SRAM}{static random access memory}
\newacronym{af}{AF}{atrial fibrillation}
\newacronym{wt}{WT}{wavelet transform}
\newacronym{cv}{CV}{cross-validation}
\newacronym[plural=SVMs,firstplural=support vector machines (SVMs)]{svm}{SVM}{support vector machine}
\newacronym{ppg}{PPG}{photoplethysmography}
\newacronym{ma}{MAs}{motion artifacts}
\newacronym{led}{LED}{light emitting diode}
\newacronym{fft}{FFT}{fast Fourier transform}
\newacronym{hr}{HR}{heart rate}
\newacronym{aae}{AAE}{Average Absolute Error}
\newacronym[plural=MCUs,firstplural=microcontroller units (MCUs)]{mcu}{MCU}{microcontroller unit}
\newacronym{loo}{LOO}{leave-one-out}
\newacronym{hrv}{HRV}{heart rate variability}
\newacronym[plural=PPVs,firstplural=positive predictive values (PPVs)]{ppv}{PPV}{positive predictive value}
\newacronym[plural=ICTs,firstplural=information and communication technologies (ICTs)]{ict}{ICT}{information and communication technology}
\newacronym{ptt}{PTT}{pulse transit time}
\newacronym{bpm}{BPM}{beats per minute}
\newacronym{ble}{BLE}{Bluetooth Low-Energy}
\newacronym[plural=ECGs,firstplural=electrocardiograms (ECGs)]{ecg}{ECG}{electrocardiogram}
\newacronym{eeg}{EEG}{electroencephalography}
\newacronym{emg}{EMG}{electromyogram}
\newacronym{iot}{IoT}{Internet of Things}

\usepackage{mfirstuc}
\MFUnocap{and}
\MFUnocap{in}
\MFUnocap{for}
\MFUnocap{a}
\MFUnocap{on}
\MFUnocap{of}
\MFUnocap{to}
\MFUnocap{vs}
\MFUnocap{versus}
\MFUnocap{the}
\MFUnocap{with}
\MFUnocap{is}
\MFUhyphentrue
\newcommand{\capitalisesection}{\capitalisewords} 
\newcommand{\capitalisesubsection}{\capitalisewords} 
\newcommand{\algorithmName}{REWARD}
\newcommand{\adaptiveAlgName}{BayeSlope}

\newcommand{\orcid}[1]{\href{https://orcid.org/#1}{\includegraphics*[width=8pt]{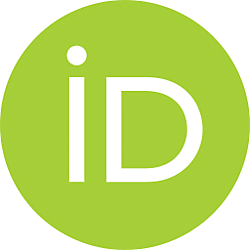}}}

\def\BibTeX{{\rm B\kern-.05em{\sc i\kern-.025em b}\kern-.08em
    T\kern-.1667em\lower.7ex\hbox{E}\kern-.125emX}}
\begin{document}
\bstctlcite{IEEEexample:BSTcontrol}%Must be the first citation in the document, so first line AFTER begin{document}
%\title{Robust and Adaptive R-Peak Detection on Low-Power Heterogeneous Monitoring Platforms}
% As alternative title for the target special issue
\title{Adaptive R-Peak Detection on Wearable ECG Sensors for High-Intensity Exercise}
%, Miguel P\'eon-Quir\'os
\author{Elisabetta De Giovanni \orcid{0000-0003-3032-5140},  Tomas Teijeiro\orcid{0000-0002-2175-7382}, Grégoire P. Millet\orcid{0000-0001-8081-4423}, and David Atienza\orcid{0000-0001-9536-4947}, \IEEEmembership{Fellow, IEEE}
\thanks{Manuscript submitted December 7, 2021. This work was supported in part by the Swiss NSF ML-Edge Project under Grant 200020\_182009, in part by the MyPreHealth Project funded by Hasler Stiftung under Grant 16073, and in part by the H2020 DeepHealth Project under Grant GA 825111.}
%, Miguel P\'eon-Quir\'os
%,  miguel.peon@epfl.ch
\thanks{Elisabetta De Giovanni, Tomas Teijeiro, and David Atienza are with the Embedded Systems Laboratory (ESL), EPFL, Lausanne, Switzerland (e-mails: elisabetta.degiovanni91@gmail.com, tomas.teijeiro@epfl.ch, david.atienza@epfl.ch). Grégoire P. Millet is with the Institute of Sport Sciences, University of Lausanne, Switzerland.}
% \thanks{Elisabetta De Giovanni is with the Embedded Systems Laboratory (ESL), EPFL, Lausanne, Switzerland (e-mail: elisabetta.degiovanni@epfl.ch). }
% \thanks{Tomas Teijeiro is with the Embedded Systems Laboratory (ESL), EPFL, Lausanne, Switzerland (e-mail: tomas.teijeiro@epfl.ch). }
% \thanks{Miguel P\'eon-Quir\'os is with the Embedded Systems Laboratory (ESL), EPFL, Lausanne, Switzerland (e-mail: miguel.peon@epfl.ch). }
% \thanks{David Atienza is with the Embedded Systems Laboratory (ESL), EPFL, Lausanne, Switzerland (e-mail: david.atienza@epfl.ch). }
}

\maketitle

% Structured abstract: Less than 250 words with objective, methods, results, conclusion, and significance.

\begin{abstract}
\textit{Objective:} Continuous monitoring of biosignals via wearable sensors has quickly expanded in the medical and wellness fields. At rest, automatic detection of vital parameters is generally accurate. However, in conditions such as high-intensity exercise, sudden physiological changes occur to the signals, compromising the robustness of standard algorithms.
\textit{Methods:} Our method, called BayeSlope, is based on unsupervised learning, Bayesian filtering, and non-linear normalization to enhance and correctly detect the R peaks according to their expected positions in the ECG. Furthermore, as BayeSlope is computationally heavy and can drain the device battery quickly, we propose an online design that adapts its robustness to sudden physiological changes, and its complexity to the heterogeneous resources of modern embedded platforms. This method combines BayeSlope with a lightweight algorithm, executed in cores with different capabilities, to reduce the energy consumption while preserving the accuracy.
\textit{Results:} BayeSlope achieves an F1 score of 99.3\% in experiments during intense cycling exercise with 20 subjects. Additionally, the online adaptive process achieves an F1 score of 99\% across five different exercise intensities, with a total energy consumption of 1.55±0.54~mJ.
\textit{Conclusion:} We propose a highly accurate and robust method, and a complete energy-efficient implementation in a modern ultra-low-power embedded platform to improve R peak detection in challenging conditions, such as during high-intensity exercise.
\textit{Significance:} The experiments show that BayeSlope outperforms a state-of-the-art algorithm up to 8.4\% in F1 score, while our online adaptive method can reach energy savings up to 38.7\% on modern heterogeneous wearable platforms.
\end{abstract}

\begin{IEEEkeywords}
Adaptive R Peak detection, Machine Learning, Edge Computing, Algorithm Robustness, Biosignal  Processing,   Heterogeneous Nodes, Energy-Accuracy Trade-Off, Ultra-Low Power Computing, Intense Exercise, Wellness
\end{IEEEkeywords}

\section{\capitalisesection{Introduction}\label{sec:ad-intro}}
In recent years, increasing healthcare costs~\cite{WHOhealthcost2018} and hospital overcrowding have pushed new technological advances to improve remote wellness monitoring, and enable early intervention and prevention~\cite{Al-khafajiy2019}.
% Increasing healthcare costs~\cite{WHOhealthcost2018} and hospital overcrowding call for new technological advances that improve remote wellness monitoring, and enable early intervention and prevention~\cite{Al-khafajiy2019}.
In addition, population aging and the resulting higher incidence of \glspl{ncd} create the need for long-term health and wellness monitoring.
For these reasons, there is an increasing demand for automatic applications working on wearable platforms that continuously and remotely monitor biosignals, such as \gls{ecg}~\cite{Kumar2018} or \gls{ppg}, and extract relevant health parameters from them. Furthermore, daily physical activity is highly recommended~\cite{Ilkka2020} to prevent \glspl{ncd}, and in particular high-intensity interval training (HIIT) are postulated as a good alternative to moderate intensity for health improvement \cite{Stamatakis2021}. As we illustrate in Section \ref{sec:ad-background}, during intense physical exercise sudden physiological changes occur, such as short RR intervals, high breathing frequency and noise from respiratory sinus arrhythmia, or sympathetic activation, amongst others~\cite{Sharma2010}. These changes induce artifacts or noise that are in general not properly detected by standard algorithms, leading to a need for using advanced computing techniques, such as machine learning and online adaptivity, to improve the robustness of the analysis. However, with the advancement of new complex algorithms comes the question of managing constrained resources in \glspl{wsn}, and the consequent toll on energy consumption. 

\begin{figure*}[tp]
	\centering
	\subfloat[Rest]{\includegraphics[width=0.45\linewidth]{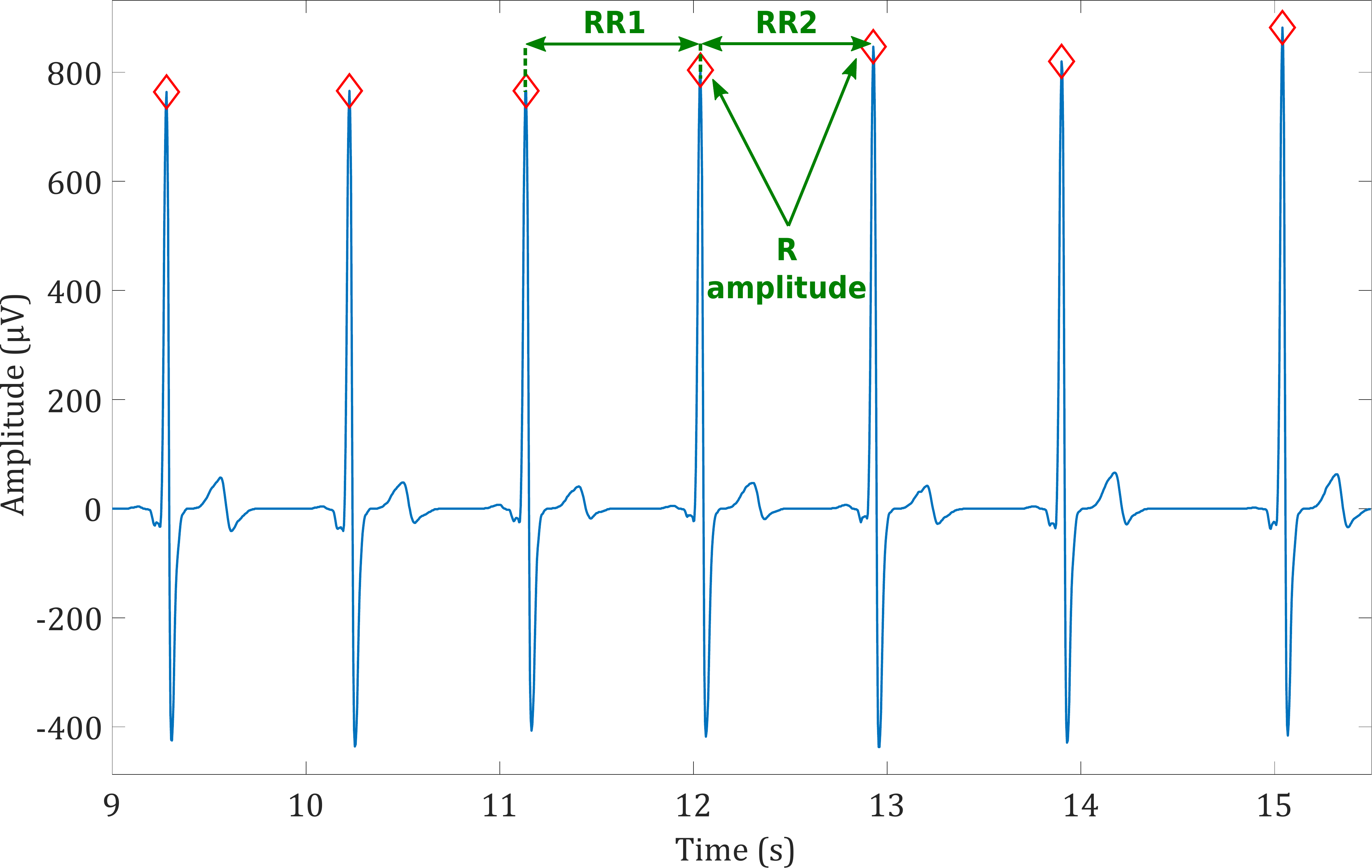} \label{fig:ad-ecg_rest}}
	\qquad
	\subfloat[Intense physical exercise]{\includegraphics[width=0.45\linewidth]{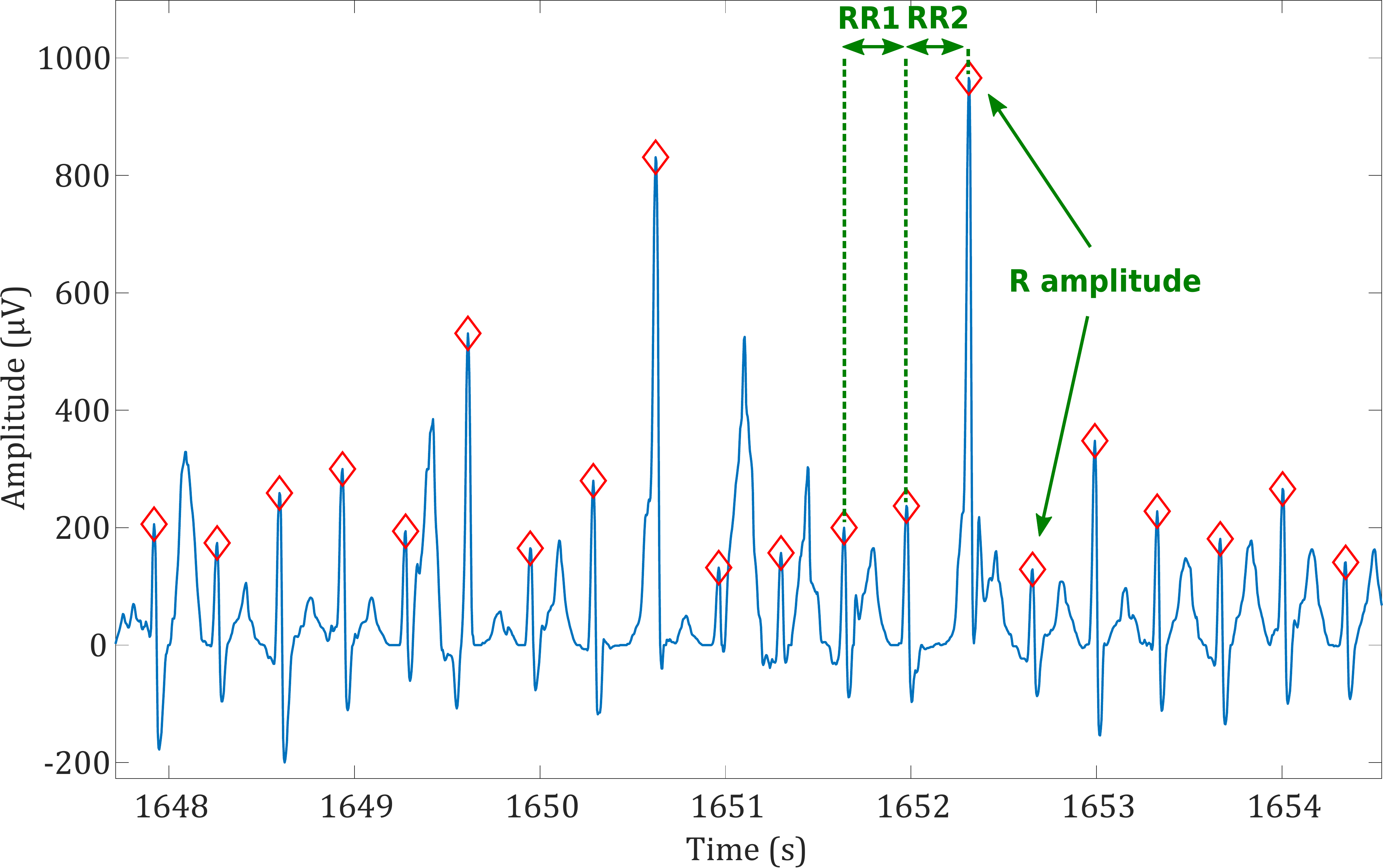} \label{fig:ad-ecg_exc4}}	
	\caption{Effects of intense physical exercise on \gls{ecg}, and, specifically, the R peak amplitudes and RR interval variability, compared to rest. The \gls{ecg} segments are extracted from the database presented in Section~\ref{sec:ad-expsetup}. Specifically, the segment in Fig.~\ref{fig:ad-ecg_rest} is extracted from the first  \SI{3}{\min} of rest of the maximal exercise test of Subject 3, starting at second nine. The segment in Fig.~\ref{fig:ad-ecg_exc4} is extracted close to exhaustion of Subject 3 during the maximal exercise test, starting at approximately \SI{27}{\min}.} 
	\label{fig:ad-ecg_changes}
\end{figure*}
In fact, the implementation of complex biomedical applications in traditional \glspl{wsn} can cause a significant draining of platform resources leading to frequent device charging~\cite{Chandrakasan2008}. Furthermore, various algorithmic optimizations implemented to lower the device energy consumption can lead to a decrease in the output accuracy of the algorithm~\cite{Zoni2020}. With the advent of modern \gls{ulp} platforms~\cite{Pullini2019} and their capabilities, the trade-off between optimizing the device resources to lower the energy consumption and maintaining a highly accurate output has become more attainable~\cite{Benatti2019,DeGiovanni2020_TETC}. Nevertheless, in the context of complex biomedical applications for \mbox{\gls{wsn}-based} wellness monitoring, the designer has to consider new challenges to achieve an optimal energy-accuracy trade-off. First, in the acquired biosignals of various pathologies or physical conditions sudden events occur, which traditional algorithms can miss or misinterpret (e.g., \gls{af} or intense physical exercise)~\cite{Kirchhof2016,Sharma2010}. For this reason, their robustness is compromised and severely affects the reliability of the wellness progress in the long term. Second, the static nature of traditional algorithms do not include the adaptive management of platform resources at run time according to the complexity of the application, which has recently become a need in \glspl{wsn} design. New methods tackle \mbox{self-aware} applications at the algorithmic level applying a \mbox{multi-layer} classification or detection system with increasing complexity~\cite{Sopic2018myo,Forooghifar2019}. Based on the confidence of the low complexity classifiers or the detection of pathological events, the algorithm decides whether to execute a more complex layer and, therefore, consumes more energy. However, these methods are targeted to traditional homogeneous platforms, and some do not consider the error in the pathological events detection. 

For the reasons above, in this work, we propose an online adaptive design of a new \gls{ecg} R peak detection algorithm for wearable systems based on machine learning, which exploits the capabilities and heterogeneity of modern \gls{ulp} platforms. In the proposed design, we introduce for the first time \adaptiveAlgName{}, a slope-based R peak detection that applies a Bayesian filter, non-linear normalization, and a clustering technique to an \gls{ecg} segment. In the literature, the use of slope-based QRS detectors has been extensive~\cite{Kohler2002,Martinez2004}. There are examples of the use of the Kalman filter for smoothed estimation of the \gls{hr}, different than R peak detection, and using multiple signals~\cite{Li2008}. However, many of these works target ambulatory monitoring. Hence, to the best of our knowledge, this is the first time that an R peak detection like \adaptiveAlgName{} is applied in the context of intense physical exercise. In fact, we test the proposed method with a dataset collected in collaboration with the Institut des sciences du sport de l'Universit\'e de Lausanne (ISSUL), where the subjects performed a maximal exercise test on a cycle ergometer till exhaustion. The outcomes of this contribution are:

\begin{itemize}
	\item We propose a new highly accurate slope-based R peak detection method, called \adaptiveAlgName{}, based on unsupervised learning. Our new R peak detection method applies a Bayesian filter and a non-linear normalization to the input \gls{ecg} signal. The combination of these signal processing techniques enhances and correctly detects the next R peak in the expected position on a peak-to-peak resolution. During high intensity exercise, \adaptiveAlgName{} outperforms the standard GQRS detector~\cite{WFDB_Moody2021} of up to \SI{8.4}{\percent} in $F_1$ score, while being comparable during low intensity exercise.
	\item We pair the newly proposed algorithm with the \algorithmName{} algorithm, presented in \cite{Orlandic2019}, which is less complex though more prone to error if sudden events occur. To ensure the adaptive nature of the design, we propose an error detection routine applied to \algorithmName{} that triggers BayeSlope if \algorithmName{} fails. 
	\item To apply adaptive management of resources at run time according to the algorithm's complexity, we implement the proposed method on an heterogeneous platform, that allows to run \adaptiveAlgName{} on a more capable core than the one where \algorithmName{} runs, which is simpler. In fact, the R peak detection step of \algorithmName{} is approximately \SI{104}{\times} less complex than \adaptiveAlgName{} when running on the same core. Hence, a simpler processor can handle it better, while a more powerful core handles better the more complex \adaptiveAlgName{}.
	\item The fully adaptive process has an $F_1 $ score of up to \SI{99.0}{\percent}, comparable to always running \adaptiveAlgName{}, which achieves an $F_1$ score up to \SI{99.3}{\percent}, across five different exercise intensities. Moreover, our proposed adaptive process is up to \SI{17.5}{\percent} more accurate compared to running only \algorithmName{}, across the five exercise intensities. Finally, the adaptive method tailored for modern heterogeneous platforms can reach energy savings up to \SI{38.7}{\percent} compared to continuously executing \adaptiveAlgName{}. Therefore, the newly proposed adaptive design is the best solution for an optimal energy-accuracy trade-off for long-term wellness monitoring with latest wearable systems. 
\end{itemize}

In Section~\ref{sec:ad-background}, we describe what occurs during intense physical exercise and the relevance of a highly accurate R peak detection in such conditions. In Section~\ref{sec:ad-methods}, we present the new R peak detection algorithm and its adaptive design. In Section~\ref{sec:ad-expsetup}, we describe the protocol of the experiments and the platform used. Finally, in Section~\ref{sec:ad-results} and Section~\ref{sec:ad-conclusion}, we present, respectively, the results and the conclusion of our analysis. 
\section{\capitalisesection{Background}\label{sec:ad-background}} 
To motivate the newly proposed method for autonomous wellness monitoring, let us consider the sudden changes occurring in the \gls{ecg} during intense physical exercise. We will focus specifically on the R peak, as it is the basis for \gls{ecg} analysis. 
% for the diagnosis of \glspl{ncd} and for the analysis of the \gls{hrv} in wellness monitoring.
Fig.~\ref{fig:ad-ecg_changes} shows two segments of \gls{ecg} acquired from a subject performing a maximal exercise test (c.f.~Section~\ref{sec:ad-expsetup}). The segments were extracted from the initial rest condition (Fig.~\ref{fig:ad-ecg_rest}) and a window of intense physical exercise, close to exhaustion (Fig.~\ref{fig:ad-ecg_exc4}). As shown in Fig.~\ref{fig:ad-ecg_exc4}, the peak-to-peak (RR) interval variability is significantly low compared to a rest condition. Moreover, the amplitude of the R peaks is highly variable. Therefore, when standard algorithms are used to detect the peaks in these conditions, such as the GQRS detector of the Physionet WFDB software package~\cite{WFDB_Moody2021} (c.f.~Section~\ref{subsec:ad-accuracy}), their robustness is compromised. In this work, we consider one R peak detection algorithm that was presented in \cite{Orlandic2019}, called \algorithmName{}, as the standard base algorithm to build on and motivate our proposed online adaptive method. \algorithmName{} can detect peaks within a window of \SI{1.75}{\second} by adapting hysteresis thresholds based on the average and maximum (or minimum if R is negative) amplitude of the peaks within a window. This method works well when the amplitude variability is limited, as shown in Section~\ref{subsec:ad-accuracy}. However, during intense physical exercise, the RR interval decreases significantly and the amplitude highly varies between one peak and the other---within \SI{1.75}{\second}, there are many peaks significantly different in amplitude---with the result that \algorithmName{} fails to detect smaller peaks. In Fig.~\ref{fig:rw_missed_peak}, we show an example of an \gls{ecg} segment extracted from the analyzed dataset and, specifically, a window where \algorithmName{} fails. 

To capture the changes occurring during various intensities of physical exercise in the wellness context, there exists a gold standard protocol where subjects perform a maximal exercise test on a cycle ergometer or a treadmill till exhaustion. The subjects wear a gas mask that measures the volume of O$_{2}$ and CO$_{2}$ (VO$_{2}$, VCO$_{2}$) inhaled and exhaled \cite{Gosselink2004}. Additionally, the protocol includes the acquisition and analysis of a single-lead \gls{ecg}, from which specific \gls{hrv} parameters can be extracted to help in the estimation of the so-called ventilatory thresholds (VT1, VT2)~\cite{Cottin2006}, and VO$_2$max~\cite{Bentley2005}. These three variables describe the cardiovascular and respiratory state during intense physical exercise. VT1 measures the hyperpnea (i.e., faster breathing) caused by the increased production of CO$_{2}$ for exercise intensities above the anaerobic threshold resulting in a non-linear increase in the ratio between ventilatory flow (VE) and VO$_{2}$. VT2 represents a phase where the hyperpnea is not enough to eliminate the CO$_{2}$, which remains constant, leading to a sharp increase of VE/VCO$_{2}$. Finally, VO$_{2}$max is the final stage where exhaustion is reached and, consequently, a maximum oxygen uptake and \gls{hr}. The determination of the ventilatory thresholds usually relies on an agreement between medical experts who evaluate the gas analysis and the \gls{hrv} parameters to find the position of the thresholds~\cite{Buchheit2007}.  

The \gls{hrv} analysis uses the RR time series of an \gls{ecg} signal to extract time and frequency domain features, which can be used for a direct estimation of VT1, VT2 and VO$_{2}$max~\cite{Cottin2006}. The current methods for this estimation are performed in post-processing with the help of medical experts and, usually, require interpolation and correction of the RR time series. The R peak detection needs to be accurate, robust and adapt at run time to sudden changes to ensure the correct comparison between ventilatory measurements and the RR time series. Therefore, we propose \adaptiveAlgName{}, a new highly accurate and robust R peak detection algorithm for wearable sensors, which is paired to \algorithmName{}. \adaptiveAlgName{} is much more complex than \algorithmName{}, and, if run continuously, it can drain the device battery. For this reason, we additionally propose a real-time strategy that automatically adapts the algorithm's complexity and the resources assigned based on the robustness of \algorithmName{}, for an enhanced energy-accuracy trade-off in latest ultra-low power autonomous wearable systems. 

 %Moreover, in future works (c.f.~Section
% %\ref{sec:future-work}
% and Appendix
% %~\ref{ch:appendix}
% ), the ventilatory threshold detection based on \gls{hrv} parameters could be performed in real-time on wearable sensors. In this case, the R peak detection needs to be energy-efficient and adapt at run time to the sudden changes that affect the \gls{ecg} during intense physical exercise~\cite{Simoons1975}. 
\section{\capitalisesection{Adaptive R peak detection in modern wearable sensors}\label{sec:ad-methods}}
One of the main problems in the context of edge computing in \glspl{wsn} is minimizing energy consumption while maximizing output accuracy. In this section, we describe our proposed method to detect R peaks from a single-lead \gls{ecg} that optimizes the energy-accuracy trade-off with a two-level adaptive method. 
\begin{figure}[tp]
	\centering
	\includegraphics[width=1\linewidth]{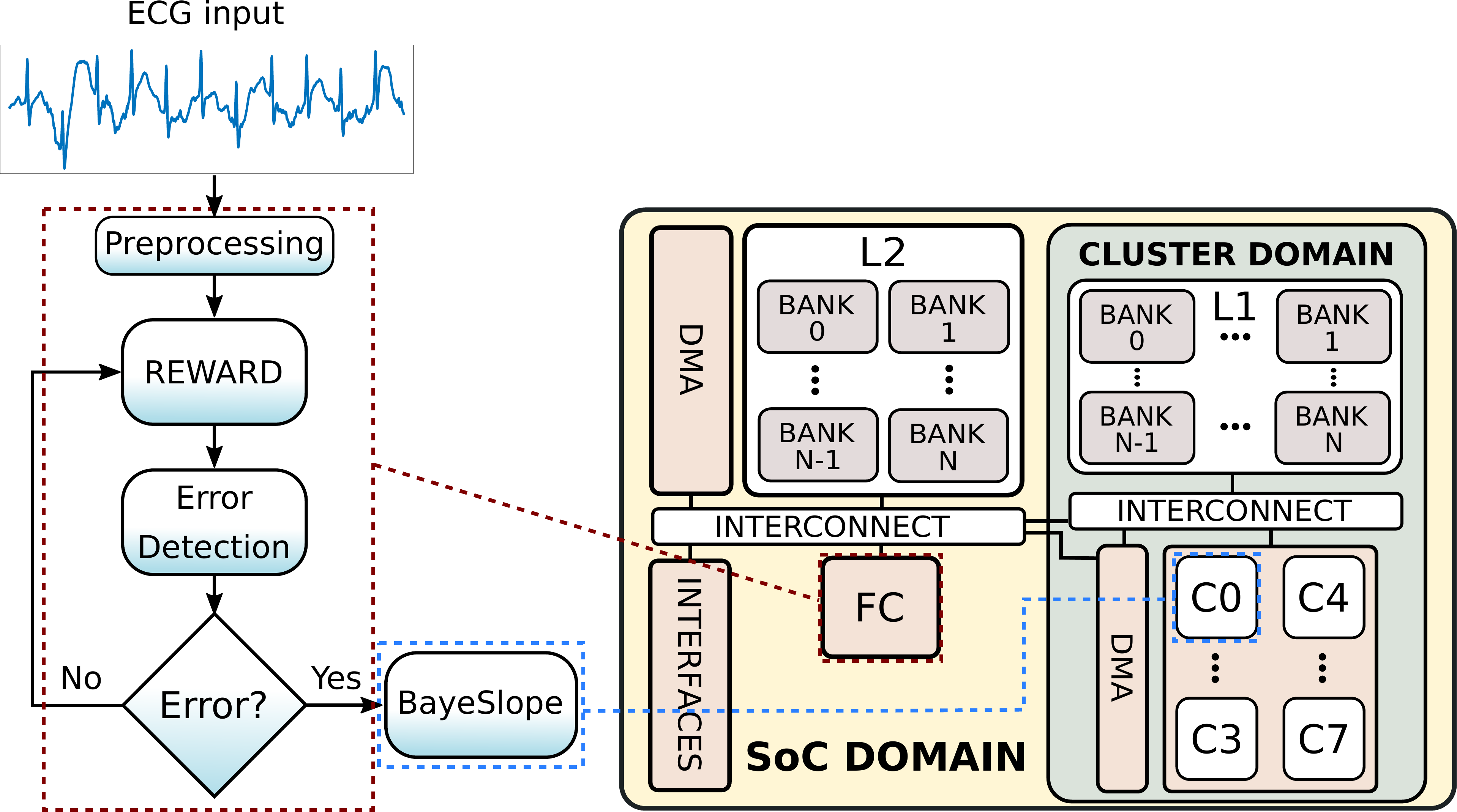}	
	\caption[Proposed online adaptive design for R peak detection and mapping on architecture]{On the left, data-flow diagram of the adaptive R peak detection algorithm with a raw \gls{ecg} input. \algorithmName{} refers to the low complexity R peak detection presented in~\cite{Orlandic2019}. \adaptiveAlgName{} is a new slope-based R peak detection algorithm presented in this work. On the right, the PULP-based~\cite{Pullini2019} architecture used for the analysis. Preprocessing, \algorithmName{}, and error detection run on the SoC domain in the fabric controller (FC), while \adaptiveAlgName{} runs on the cluster domain, in one core of the cluster (CL) of eight cores.}
	\label{fig:method_steps}
\end{figure}
Fig.~\ref{fig:method_steps} shows the data-flow diagram of the full process and the architecture where the algorithm is implemented~\cite{Pullini2019}. The two-level adaptivity consists in the robustness and complexity of two different R peak detection algorithms, namely \algorithmName{}~\cite{Orlandic2019} and the newly proposed algorithm, \adaptiveAlgName{}. As described in Section~\ref{sec:ad-background}, \algorithmName{} uses hysteresis thresholds that are adapted to each window of \SI{1.75}{\second}. However, this window resolution is too small to capture peak-to-peak sudden changes. For this reason, we introduce \adaptiveAlgName{} that implements peak normalization through a generalized logistic function and a Bayesian filter to enhance and compute the expected position of the next R peak. Once the peaks are normalized and enhanced, the method applies a k-means clustering to divide the \gls{ecg} samples in two clusters represented by two centroids, one corresponding to the baseline and one to the R peak. The first level of adaptivity consists in adapting the robustness of the algorithm, as shown in the data-flow of Fig.~\ref{fig:method_steps}. The output of \algorithmName{} is fed to an error detection method that checks when \algorithmName{} fails to properly detect R peaks and, in this case, triggers the more accurate and robust \adaptiveAlgName{}. The second level of adaptivity consists in adapting the resources to the complexity of the method. In fact, \adaptiveAlgName{} is a more complex algorithm, hence, it benefits from execution on the cluster of cores in the platform, which includes eight cores with higher \gls{ipc} and floating point units. Secondly, the main core, which is a simpler core, can handle the less complex \algorithmName{} in a more energy-efficient way. In the next sections, we describe first the different blocks shown in Fig.~\ref{fig:method_steps}. Then, the higher-level design within the heterogeneous platform is presented. 

\subsection{\capitalisesubsection{Preprocessing, \algorithmName{} and error detection}\label{ad-subsec:preproc_rw_errdet}}
A standard R peak detection algorithm requires several steps of preprocessing of the \gls{ecg} input signal. In this case, the input is a single-lead \gls{ecg} where a \gls{mf} is applied to remove the baseline and high frequency noise~\cite{Braojos2012}. Then, the signal is enhanced by applying the \gls{relen} method, which amplifies the most dominant peaks~\cite{Orlandic2019}. This preprocessing method is part of the \algorithmName{} algorithm presented in~\cite{Orlandic2019}.
\begin{figure}[tp]
	\centering
	\includegraphics[width=1\linewidth]{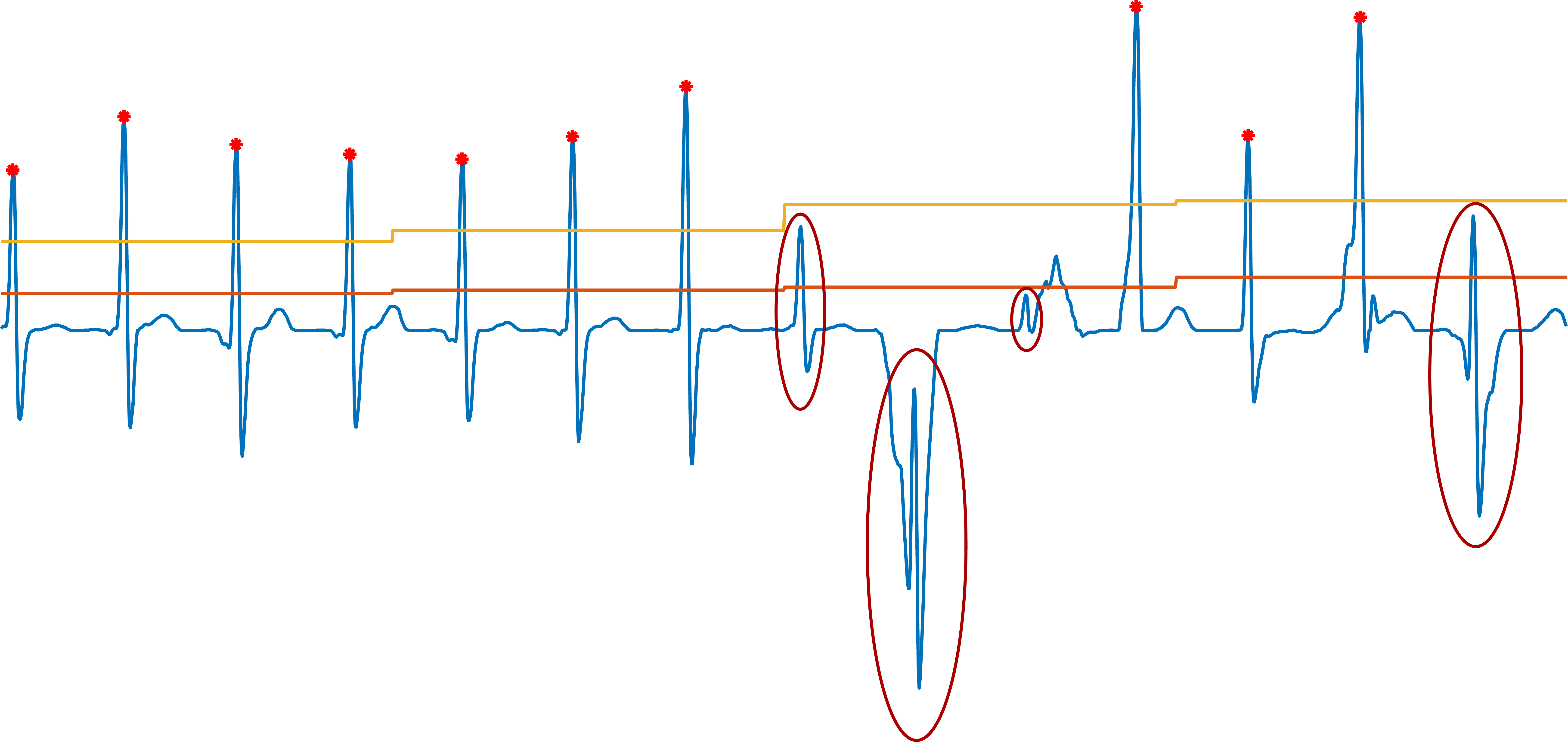}	
	\caption[Missed peaks by \algorithmName{} when sudden changes in ECG occur]{Missed peaks in \algorithmName{} R peak detection using hysteresis thresholds (in orange and yellow) based on the \gls{ecg} window morphology. The segment was extracted from Subject~3 of the used dataset (c.f.,~Section~\ref{subsec:ad-database}).}
	\label{fig:rw_missed_peak}
\end{figure}
The second part of the algorithm searches for the R peak in a window of  \SI{1.75}{\second} using hysteresis thresholds based on the \gls{ecg} morphology within the window. However, during intense physical exercise, the interval between two R peaks (i.e., RR interval) decreases significantly and sudden changes in amplitude occur. Therefore, within a window of analysis, many peaks can be missed, as shown in Fig.~\ref{fig:rw_missed_peak}. Moreover, right after exhaustion during a maximal exercise test, there can be an increase in T wave---the wave after the R peak that represents the repolarization of the heart ventricles---often significantly more dominant than the R peak itself, and a decrease of the RT interval. In these conditions, \algorithmName{} fails in detecting very small peaks as the hysteresis thresholds are skewed by the higher amplitude variability of the peaks within the window. However, it performs extremely well if these events do not occur as demonstrated in~\cite{Orlandic2019}. %the P wave---the wave before the R peak that represents the depolarization of the atria---

For this reason, we propose a method to identify errors in the R peak detection within a window of \SI{1.75}{\second} analyzing the distribution of the ratio $\frac{RR(n)}{RR(n-1)}$, where $n=0,1,2...$, of all the data acquired. This distribution has been computed offline using the results of \adaptiveAlgName{}, since it is the most accurate (c.f., Section~\ref{sec:ad-results}). However, to avoid data snooping, for each subject, the RR ratio distribution is computed with a \gls{loo} strategy, in which the analyzed subject is not included in the distribution. The RR ratio can capture sudden changes with a three-peak resolution, such as missing peaks, additional wrong peaks (e.g., T wave), and highly noisy signal segments. 

\begin{figure}[tp]
	\centering
	\includegraphics[width=1\linewidth]{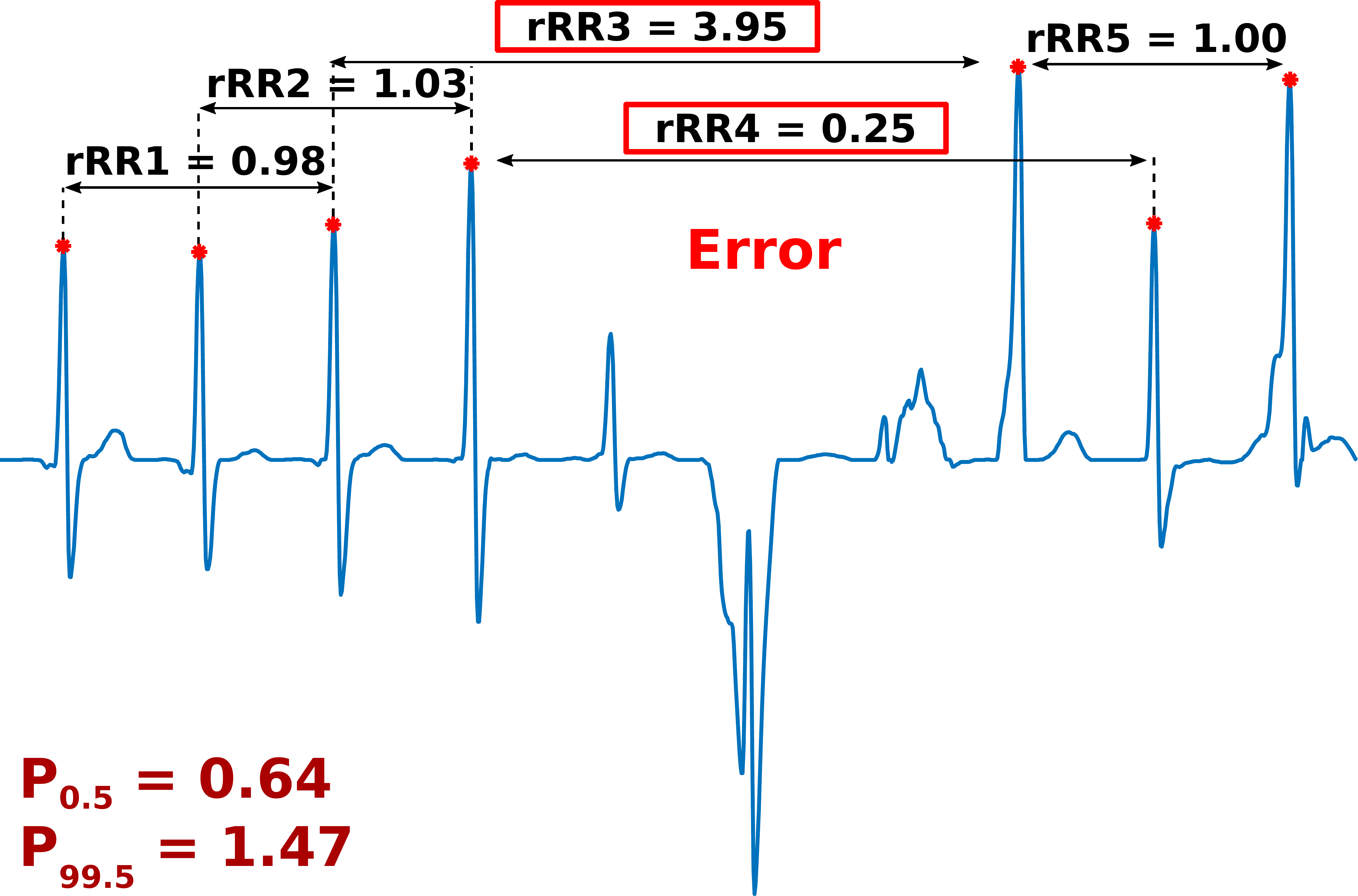}	
	\caption[Result of error detection on an example ECG]{Result of error detection on example \gls{ecg} extracted from Subject~3 of the dataset used (c.f.,~Section~\ref{subsec:ad-database}). The values of the RR ratio are computed on a resolution of three R peaks. Considering the percentile thresholds for the analyzed subject (bottom left), the method can detect an error where \algorithmName{} fails (in the red boxes). }
	\label{fig:rw_missed_peak_errdet}
\end{figure}

First, the method computes offline the RR intervals and the corresponding RR ratio sequence used for the distribution from all the subjects, except the one that is being analyzed. Then, for each subject, if at least one value of RR ratio computed within each window falls in the tails of the distribution (below the 0.5 or above the 99.5 percentiles of the RR ratio distribution, respectively), the algorithm detects an error. This is performed in the online phase of the error detection applied to the output R peaks of \algorithmName{}. 
\begin{figure}[tp]
	\centering
	\includegraphics[width=0.9\linewidth]{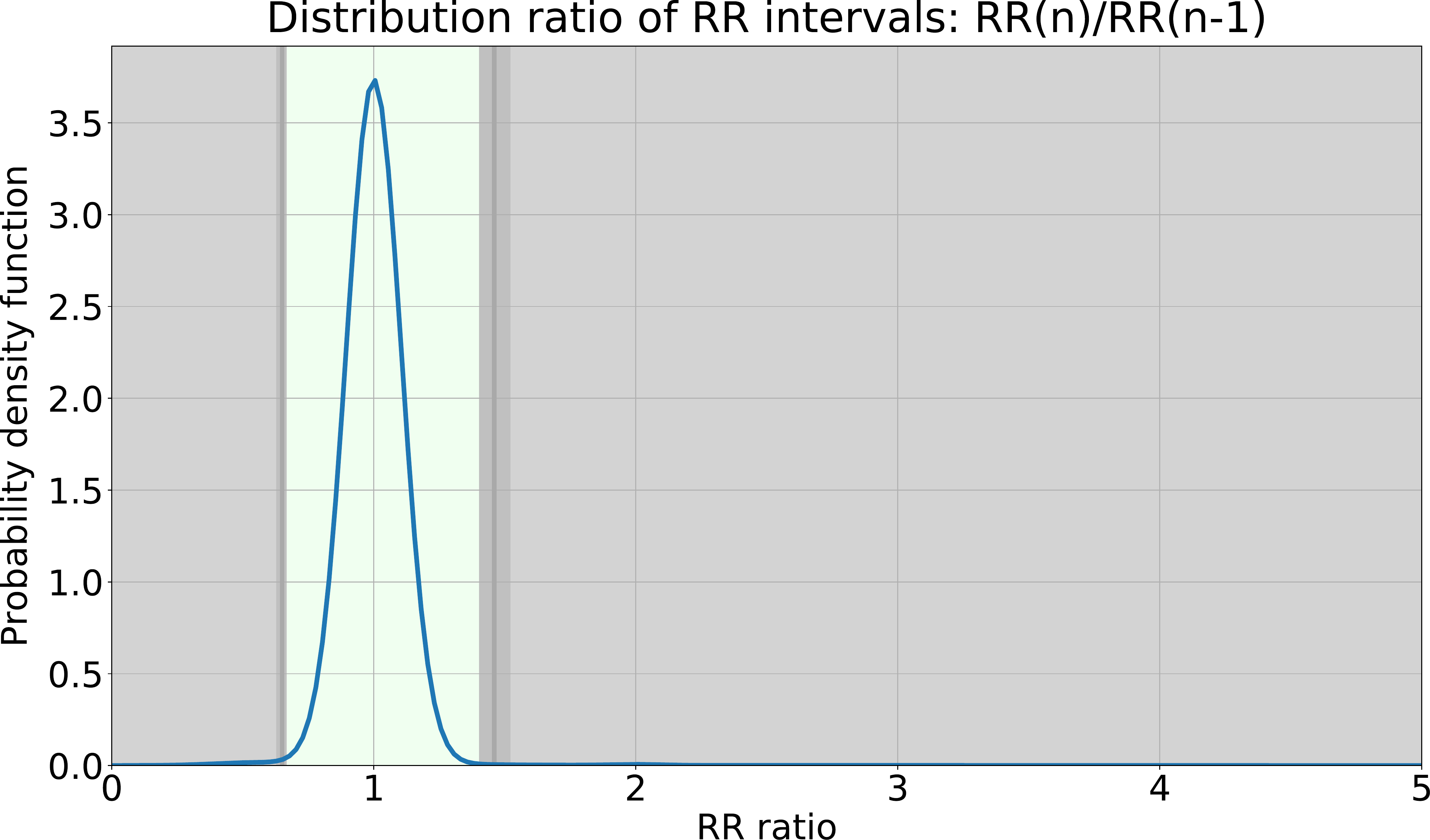}	
	\caption[Distribution of RR ratio for the error detection step in online adaptive design]{RR ratio distribution for the full dataset acquired for this analysis. The shadowed grey areas represent the range of RR ratio for which an error is detected, while the light green area represents the range of RR ratio for which there is no error in the R peaks. 
	% ==== REMOVE THE FOLLOWING PARAGRAPH IF YOU USE distributionRR_ratio.pdf OR PUT IT IF YOU USE distributionRR_ratio_rangeThs.pdf ====
	Moreover, the darker grey areas and the vertical lines represent the full range of percentile thresholds for all the subjects in the used dataset (c.f.,~Section~\ref{subsec:ad-database}).
	% ==== PARAGRAPH ====
	}
	\label{fig:distributionRR}
\end{figure}
Fig.~\ref{fig:distributionRR} shows the distribution considering all the subjects analyzed in this work. We report the overall distribution for convenience, but note that it is not the one used in our proposed online error detection. Moreover, the right tail is longer than reported on the figure, as it is redundant. In fact, the percentile thresholds with the \gls{loo} strategy are:
\begin{equation}
	P_{0.5} = 0.65\pm0.02; \\
	P_{99.5} = 1.46\pm0.06;
	\label{eq:percentilesLOO}
\end{equation}
Even though the distributions are different, the standard deviation among the subjects is quite small, suggesting that the RR ratio has a low inter-patient variability. On the other hand, the range of the distribution suggests a high intra-patient variability. 
Considering these values of thresholds, Fig.~\ref{fig:distributionRR} shows shadowed areas in grey, which represent the range of RR ratio for which an error is detected.
%% ==== PUT THE FOLLOWING LINE IF YOU USE distributionRR_ratio.pdf OR REMOVE IT IF YOU USE distributionRR_ratio_rangeThs.pdf (IN THIS CASE REMOVE LINE "in the worst case with $(P_{0.5},P_{99.5}) = (0.67,1.40)$ ====
%in the worst case with $(P_{0.5},P_{99.5}) = (0.67,1.40)$. 
%% ==== LINE ====
Moreover, this figure shows a light green area that represents the range of RR ratio for which no error in the R peaks exists.
% ==== REMOVE THE FOLLOWING PARAGRAPH IF YOU USE distributionRR_ratio.pdf OR PUT IT IF YOU USE distributionRR_ratio_rangeThs.pdf (IN THIS CASE REMOVE LINE "in the worst case with $(P_{0.5},P_{99.5}) = (0.67,1.40)$ ====
Finally, the darker grey areas and the vertical lines represent the full range of percentile thresholds reported in~(\ref{eq:percentilesLOO}). 
% ==== PARAGRAPH ====
Therefore, if we consider the \gls{ecg} example in Fig.~\ref{fig:rw_missed_peak}, the error detection results are shown in Fig.~\ref{fig:rw_missed_peak_errdet}, where the values of the RR ratio over the segment are reported. Considering the percentile thresholds $P_{0.5} = 0.64$ and $P_{99.5} = 1.47$ for the analyzed subject, the method can detect an error where \algorithmName{} fails. The last peak in Fig.~\ref{fig:rw_missed_peak} where there is an error would be detected in the next window.
It is worth mentioning that there are genuine physiological events, such as ectopic heartbeats, that can cause a variation in the RR ratio in the tails of the distribution, and therefore they will be detected as "errors". These events will lead to trigger the \adaptiveAlgName{} algorithm, but it does not mean that they will be ignored or undetected, as long as they fit the detection conditions of \adaptiveAlgName{}.

\subsection{\capitalisesubsection{\adaptiveAlgName{}: adaptive slope-based R peak detection}\label{ad-subsec:bayeslope}}
\begin{algorithm}
	\caption{\adaptiveAlgName{} R peak detection}
	\label{alg:bayeslope}
    \algrenewcommand{\alglinenumber}[1]{#1}
    \scriptsize
	\begin{algorithmic}[1]
        \State \textbf{Input}: windows of RelEn signal, $s$ (\SI{}{\micro\volt})
		\State \textbf{Output}: R peaks, $r$
		\State Initialize centroids: $hcentr =$ percentile(diff($s$),$99$) and $lcentr = 1$ \label{alg:adcl_centr_init}
		\State $min\_rr\_dist = 240$ ms; $max\_qrs\_dur = 140$ ms; \Comment{Constant parameters}\label{alg:adcl_params1_init}
		\State Initialize: $mu = 75$ bpm; $sd = 100$ ms; $zeroctr = 0$; $qrs\_init = 0$; $label = 0$; $in\_qrs =$ false;\label{alg:adcl_params2_init}
		\For{$i = 2,...$length($s$)}
			\State $s2[i] = s[i] - s[i-1]$;    \Comment{Derivative approximation}\label{alg:adcl_diff}
			\State $x =$ abs($s2[i]$);\label{alg:adcl_absdiff}
			\State $bf[i] =$ gaussian($i-last\_peak, mu, sd$);   \Comment{Bayesian filter}\label{alg:adcl_gauss}
			\State $bt[i] =$ genlogfun($x,param\_logfun$);   \Comment{Sigmoid normalization}\label{alg:adcl_logfun}
			\State $st[i] =$ max($x, bt[i]*bf[i]$); \Comment{Normalize signal}\label{alg:adcl_norm}
			\State Update $hcentr$ and $lcentr$ each as the new mean of their cluster\label{alg:adcl_kmeans} 
			\If{$in\_qrs$} \Comment{Peak search}\label{alg:adcl_peaksearch_start}
				\If{$label = 0$}\label{alg:adcl_wait_start}
					\State $zeroctr += 1$;
				\Else
					\State $zeroctr = 0$;
				\EndIf
				\If{$zeroctr = 0$ OR $i-qrs\_init > max\_qrs\_dur$}\label{alg:adcl_wait_end}
					\State $max\_min\_slope =$ argmaxmin$(st * sign(s2))$;\label{alg:adcl_start_inqrs_pksearch}
					\State Search for $new\_peak$ within $max\_min\_slope$
					\State $r[i] = new\_peak$;\label{alg:adcl_end_inqrs_pksearch}
				\EndIf
			\Else
				\If{$label = 1$ AND $i > last\_peak + min\_rr\_dist$}\label{alg:adcl_inqrs_cond}
					\State $in\_qrs =$ true;
					\State $qrs\_init = 1$;
				\EndIf
			\EndIf\label{alg:adcl_peaksearch_end}
		\EndFor		
	\end{algorithmic}
\end{algorithm} 
Once an error is detected, a more accurate adaptive R peak detection, \adaptiveAlgName{}, is triggered. This newly proposed method applies a non-linear normalization of the signal and a Bayesian filter to enhance high slope areas, which are assumed to belong to the QRS complex, and correctly detect the next peak in the expected position, which is based on the current \gls{hr}. To distinguish low and high slope areas, the approach relies on a clustering method based on K-means.

Algorithm~\ref{alg:bayeslope} describes the main steps of \adaptiveAlgName{}. The method takes as input the \gls{relen} signal window, $s$, and it outputs the vector of R peaks detected. The algorithm is derivative-based, and considers two clusters that represent the high and low slope areas of the signal. Then, the two centroids are initialized beforehand, as shown in Line~\ref{alg:adcl_centr_init}, where $hcentr$ is the 99$^{th}$ percentile of the derivative of $s$ and $lcentr = 1$. When a new sample is assigned to a cluster it is labeled as 1, if belonging to $hcentr$ cluster, or 0, if belonging to $lcentr$ cluster. Two windows of \SI{1.75}{\second} are used for the $hcentr$ initialization to account for enough peaks even at rest and avoid errors due to signal noise. Then, the algorithm initializes all the other parameters needed, constant and varying, in Lines~\mbox{\ref{alg:adcl_params1_init}--\ref{alg:adcl_params2_init}}. The values for these parameters were chosen based on physiological information but were not optimized in order to avoid loss of generality.

\begin{figure}[tp]
	\centering
	\includegraphics[width=1\linewidth]{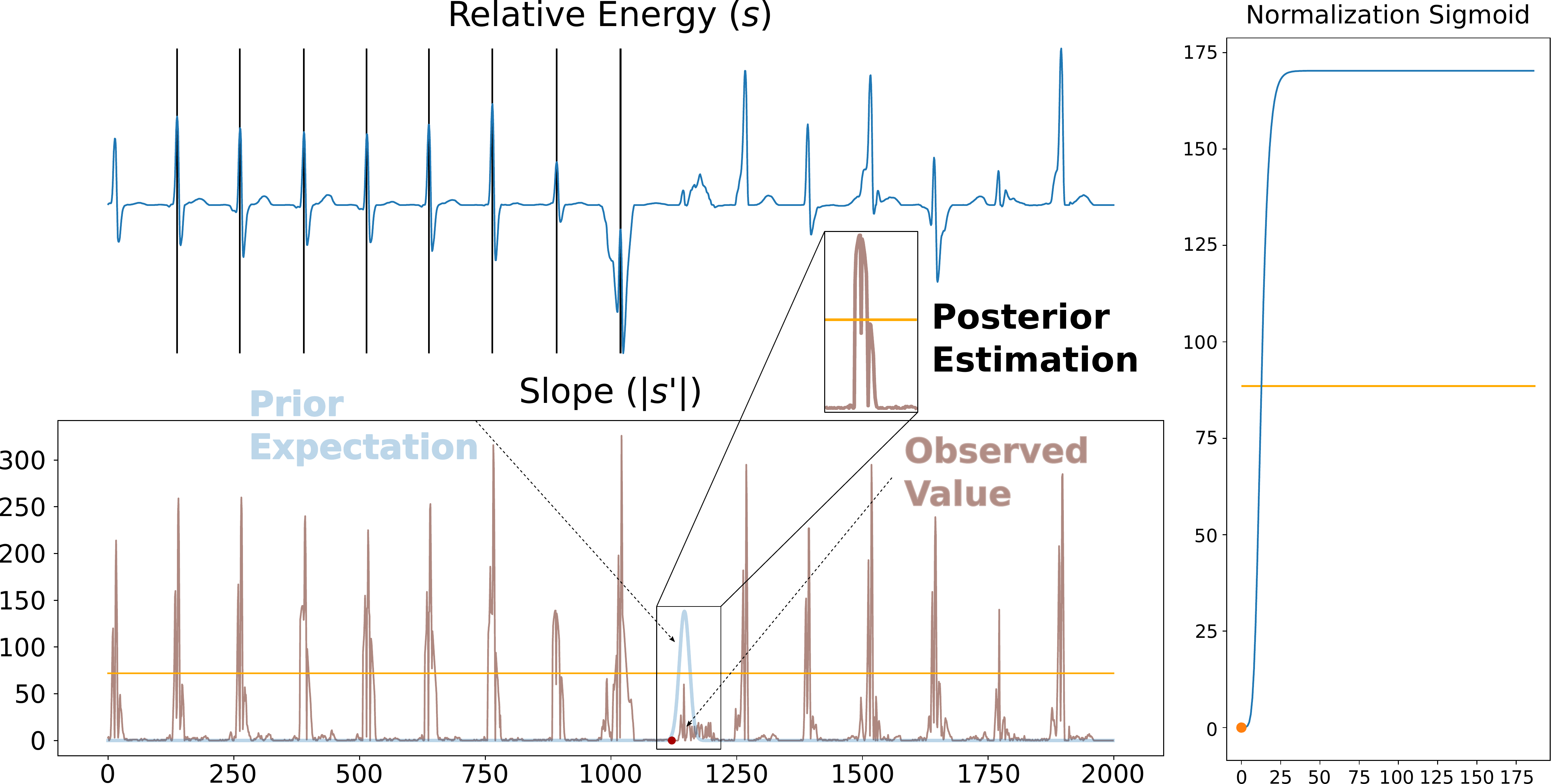}	
	\caption[Peak normalization in proposed \adaptiveAlgName{} algorithm]{Peak normalization in the expected location through bayesian filter (i.e., prior expectation) and generalized logistic function. Once the normalization is applied, the observed value is transformed into the posterior estimation, shown in the rectangle}
	\label{fig:peak_norm}
\end{figure}

The main process starts by considering the derivative of $s$ and computing its absolute value, $x$, in Lines~\mbox{\ref{alg:adcl_diff}--\ref{alg:adcl_absdiff}}. We apply this initial transformation to enhance the maximum and minimum slopes of the original signal $s$. This choice follows the assumption that the R peak is located in general within the maximum upward and downward deflections within an \gls{ecg} signal. Next, the method computes the Bayesian filter (Line~\ref{alg:adcl_gauss}), which is a Gaussian centered on the expected peak, $mu$, with standard deviation $sd$, two parameters computed based on the last five peaks. Then, in Line~\ref{alg:adcl_logfun}, the algorithm computes the generalized logistic function~\cite{RICHARDS1959} with input $x$ and its parameters computed based on the last $hcentr$ and $lcentr$. The sigmoid varies between 0 and the value of the higher k-means centroid, $hcentr$. The sigmoid and the Bayesian filter are used to normalize the peak or, specifically, to increase the amplitude of expected small peaks, as shown in Line~\ref{alg:adcl_norm} and Fig.~\ref{fig:peak_norm}. If the analyzed sample does not reach the computed threshold, the function does not increase its value. When the input is approximately double the value of the lowest centroid, $st$ (in Algorithm~\ref{alg:bayeslope}) reaches the threshold between the $lcentr$ and $hcentr$. In Fig.~\ref{fig:peak_norm}, the expected location of the peak (i.e., the prior expectation) is depicted with the Gaussian centered on it. In this case, the original peak (i.e., observed value) in $x$ ($|s'|$) is small, and it will be enhanced by the Gaussian multiplied by the sigmoid function, $st$. This situation occurs at the values where $st$ exceeds the threshold (in orange), with the result shown in the posterior estimation rectangle. 

Once the signal is normalized, the algorithm starts a peak search within a QRS complex---the \gls{ecg} main wave---in Lines~\mbox{\ref{alg:adcl_peaksearch_start}--\ref{alg:adcl_peaksearch_end}}. The distance between QRS complexes must be more than $min\_rr\_dist$, according to standard physiological characteristics and the sample that starts the QRS complex ($qrs\_init$ in Line~\ref{alg:adcl_inqrs_cond}) must belong to the cluster represented by $hcentr$ (i.e., $label = 1$). Within the QRS complex, the algorithm waits till it reaches its maximum duration according to physiology ($max\_qrs\_dur$) or for enough samples ($zeroctr = 30$) labeled 0 that represent the end of the QRS complex (Lines~\mbox{\ref{alg:adcl_wait_start}--\ref{alg:adcl_wait_end}}). Once within this interval (Lines~\mbox{\ref{alg:adcl_start_inqrs_pksearch}--\ref{alg:adcl_end_inqrs_pksearch}}), the algorithm computes the maximum and minimum of the function $st*sign(s2)$ representing the maximum upslope and downslope of the original signal. The sign function is used in case these values fall in the Q, S or T wave, which are not distinguished if only $st$ is used, as it is positive by definition. Finally, the $new\_peak$ is found and stored in the vector $r$.

\subsection{\capitalisesubsection{Adaptive design in modern heterogeneous platforms}\label{ad-subsec:adapt_platform}}
As shown in Fig.~\ref{fig:method_steps}, the modules of the algorithm run in different cores of the wearable computing architecture according to the complexity of the corresponding module. The wearable architecture used in this work is based on one of the evolutions of the open-source PULP platform~\cite{PULPSDK}, called Mr.Wolf~\cite{Pullini2019}. The PULP structure consists of a main streamlined processor, the \gls{fc}, and an 8-core parallel compute \gls{cl}. Moreover, PULP includes a \gls{dma} that can transfer data to a multi-banked \SI{512}{\kibi\byte} L2 memory during acquisition time or from L2 to a shared multi-banked \SI{64}{\kibi\byte} L1 memory, which has a single-cycle latency in the cluster side. Both \gls{fc} and \gls{cl} are power-gated while the \gls{dma} fills the required L2 memory bank during sample acquisition. The \gls{fc} is clock-gated when the \gls{cl} is active, and each of the cores in the \gls{cl} can be independently clock-gated to reduce dynamic power. Mr.Wolf includes a core for the \gls{fc} (Zero-riscy) that is simpler than the RI5CY cores of the \gls{cl}, but it has a lower IPC. On the other hand, the cores of the \gls{cl} have more capabilities~\cite{RISCV-IoT}. Therefore, this work considers the Mr.Wolf architecture by selectively using the \gls{fc} and one core of the \gls{cl}. 

Considering this design, the modules of preprocessing (\gls{mf}), \algorithmName{} (which includes \gls{relen} and R peak detection via hysteresis thresholds), and error detection run in the \gls{fc}. \algorithmName{} is a very lightweight integer-based algorithm, as demonstrated in~\cite{Orlandic2019}. In a preliminary analysis, considering the dataset (c.f.,~Section~\ref{subsec:ad-database}), we performed a test executing the R peak detection step of \algorithmName{} on the \gls{fc} and on one core of \gls{cl}. The algorithm executed on \gls{cl} is \SI{1.23}{\times} slower (in terms of execution time) and consumes \SI{1.35}{\times} more energy. Therefore, \algorithmName{} benefits from running on the \gls{fc}, which is a simpler core, clocked at a higher frequency (\SI{170}{\mega\hertz} vs. \SI{110}{\mega\hertz} of the \gls{cl}). On the contrary, \adaptiveAlgName{} is about \SI{100}{\times} more complex than \algorithmName{}, hence, it benefits from running on a more advanced core with higher \gls{ipc}. This helps to meet real-time constraints and limits the amount of time the system is active. Additionally, the Gaussian and the generalized logistic function of \adaptiveAlgName{} are implemented in floating-point. Since the \gls{fc} does not have a floating-point unit, \adaptiveAlgName{} should be converted to fixed-point representation. 
% \footnote{Another solution would be to simulate floating-point arithmetic in SW, which is slower and more energy-consuming.}
Therefore, we adapted these functions of \adaptiveAlgName{} to employ fixed-point arithmetic, using a 32-bit representation with 1 sign bit, 15 integer bits, and 16 decimal bits. The results reveal that during the clustering step the algorithm quickly reaches the maximum range representable (i.e., approximately within \SI{15}{\second} of signal processing), with a consequent drop in accuracy. In contrast, this does not occur in the 32-bit floating-point representation as the maximum range is reached after approximately \SI{27}{\hour} of signal processing. Therefore, we decided to implement \adaptiveAlgName{} on one core of the \gls{cl} (RI5CY), which has a floating point unit and higher IPC. 

After the signal filtering and \algorithmName{} running on the \gls{fc}, the error detection (also running on the \gls{fc}) checks the accuracy of the R peaks output. If an error is detected, the \gls{dma} transfers the necessary buffer of data from L2 to L1 ready for the core in the \gls{cl}. Conversely, the \gls{fc} is clock-gated. Since \adaptiveAlgName{} needs an initialization of the R peaks of two windows of \SI{1.75}{\second}, the previous window error needs to be checked. If the error in the previous window is 0, then the \gls{dma} transfers two windows, otherwise it transfers only one. This is an optimization applied in case \algorithmName{} fails more frequently and to avoid recomputing twice the same window. \adaptiveAlgName{} runs on the \gls{cl} while the \gls{fc} is clock-gated. The final output is the combination of correct R peaks from \algorithmName{} and \adaptiveAlgName{}. The full code for the adaptive R peak detection has been published as open-source software\footnote{\href{https://c4science.ch/source/adaptive_rpeak_det_public}{https://c4science.ch/source/adaptive\_rpeak\_det\_public}}. 

\clearpage %Problems in the layout with the next section. This should be removed eventually
\section{\capitalisesection{Experimental setup}\label{sec:ad-expsetup}}
\subsection{\capitalisesubsection{Database acquisition protocol}\label{subsec:ad-database}}
\begin{figure}[tp]
	\centering
	\includegraphics[width=1\linewidth]{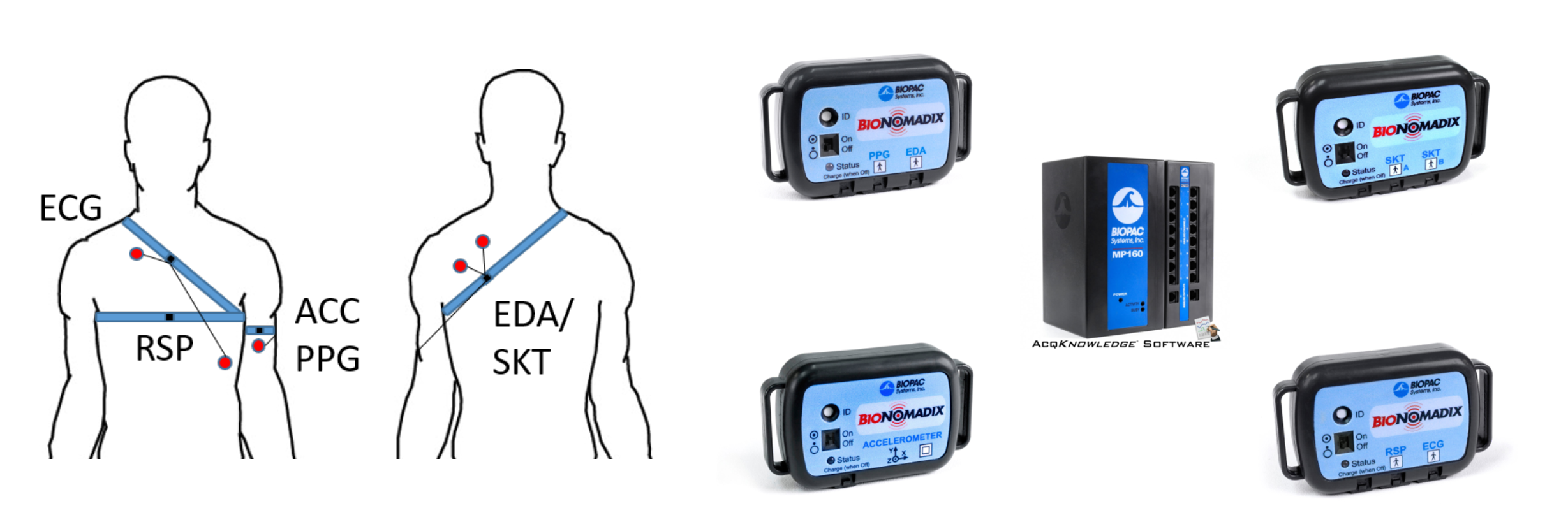}	
	\caption[Sensors positioning for acquisition of incremental exercise stress test]{Sketch of the BIOPAC~\cite{Biopac} sensors positioning (left) during the experiment and the sensors themselves (right)}
	\label{fig:protocol_exp}
\end{figure}
The database was acquired considering 22 subjects performing an incremental test to exhaustion on a cycle ergometer for an average of 30 minutes each until VO$_2$max was reached, plus at least 1 minute post-exercise. The power of the cycle ergometer was increased every \SI{3}{\min} by \SI{30}{\watt}, after initial \SI{3}{\min} of rest. Moreover, a three-minute recovery period was recorded right after exhaustion. A single-lead \gls{ecg} sampled at \SI{500}{\hertz} was acquired using the BIOPAC system~\cite{Biopac}, together with other biosignals and oxygen uptake measurements that were not used for this work. Fig.~\ref{fig:protocol_exp} shows a sketch of the positioning of biosignals sensors and the equipment used. The protocol was ethically approved by the Commission Cantonale (VD) d'Ethique de la Recherche sur l'Etre Humain (CER-VD), with reference 2016-00308, on 01/03/2018. For the experiments, the \gls{ecg} was downsampled to \SI{250}{\hertz} since \algorithmName{} was validated only for this frequency in~\cite{Orlandic2019}. Two of the 22 subjects were discarded because one did not complete the protocol and for the second one the majority of the recording was corrupted. Therefore, the statistics and analysis were performed on 20 subjects. 
\begin{figure}[tp]
	\centering
	\includegraphics[width=1\linewidth]{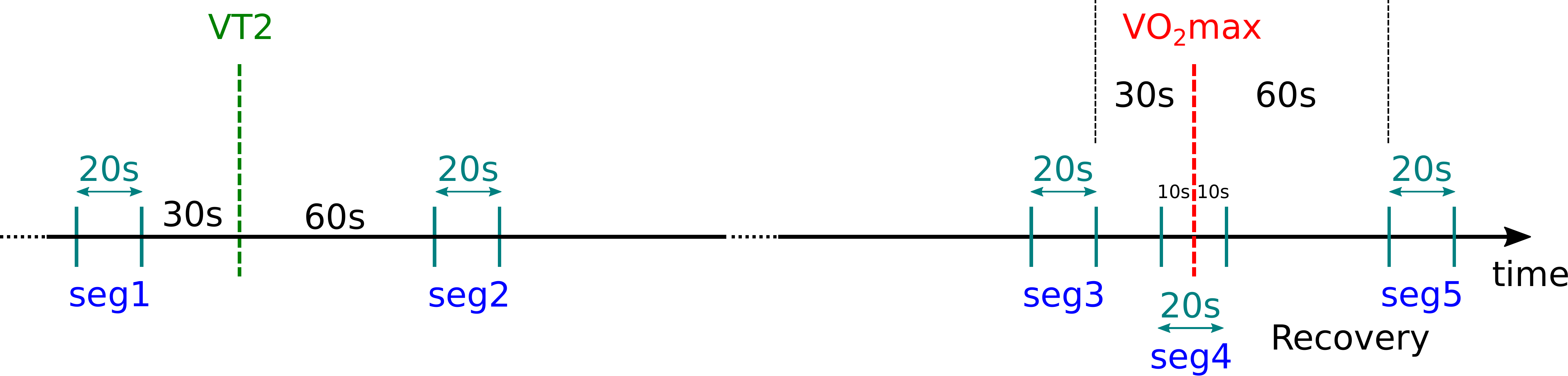}	
	\caption{Position in time through the maximal exercise test of the five 20-second segments extracted from the full \gls{ecg} of each subject. The numbers of the segments are sorted in time in ascending order. The first segment was extracted \SI{30}{\second} before VT2 and corresponds to heavy intensity; the second \SI{60}{\second} after VT2 (severe intensity); the third \SI{30}{\second} before VO$_2$max (highly severe intensity up to exhaustion); the fourth at the moment of exhaustion (centered in VO$_2$max); the fifth \SI{60}{\second} post-exercise, i.e. during the recovery after exhaustion.}
	\label{fig:segments_timeline}
\end{figure}
Next, five 20-second segments were extracted from the full \gls{ecg} of each subject to be manually annotated by medical experts. These segments were chosen based on the different phases of the maximal exercise test, namely, considering higher intensities of exercises where it is more likely that sudden changes occur. Then, these segments were extracted and reported in the order shown in Fig.~\ref{fig:segments_timeline}: the first, \SI{30}{\second} before VT2; the second, \SI{60}{\second} after VT2; the third, \SI{30}{\second} before VO$_2$max (exhaustion); the fourth, at the moment of exhaustion (centered in VO$_2$max); finally, the fifth, \SI{60}{\second} after VO$_2$max, i.e. during the recovery after exhaustion. The segments at rest were ignored since \algorithmName{} performs very well in this condition, and there is no need to run \adaptiveAlgName{}. Moreover, also the segments near VT1 are not considered as they represent lower intensities of exercise for which the performance of \algorithmName{} is satisfactory. Only one out of 100 segments was not extracted and annotated (subject~9, segment during recovery). In fact, the recording for this subject was stopped right after exhaustion (instead of after three minutes of recovery expected by the protocol) and it was not possible to have a 20-second segment \SI{60}{\second} after VO$_2$max. The input segments to the peak detection were extracted considering the \SI{20}{\second} given to the experts and going backward of  $\SI{0.6}{\second}+\SI{0.95}{\second}+\SI{1.75}{\second}$, which represents, respectively, the initial delay of the \gls{mf}, the initial delay of \gls{relen}, and one additional window of analysis for BayeSlope initialization, and going forward another \SI{1.75}{\second} to avoid missing the last peaks. Therefore, each segment is approximately \SI{25}{\second} long. The accuracy of the R peak detection is measured according to the standard tolerance of \SI{150}{\milli\second} between the detected peak and the manually annotated one~\cite{AAMI2008}. Moreover, we also report for each subject the mean and standard deviation of the time difference between the two. We compare first the accuracy of \adaptiveAlgName{} against the standard GQRS detector~\cite{WFDB_Moody2021}. Then, we compare the accuracy of the following three designs:
\begin{enumerate}
	\item preprocessing (\gls{mf}) and always running \algorithmName{} (\gls{relen} + peak detection);
	\item preprocessing (\gls{mf} + \gls{relen}) and always running the newly proposed \adaptiveAlgName{};
	\item our proposed adaptive design including preprocessing (\gls{mf}), \algorithmName{} (\gls{relen} + peak detection), error detection and running \adaptiveAlgName{} only when \algorithmName{} fails. 
\end{enumerate}
 All the segments used in the experimental validation, as well as the manual annotations, have been published as an open dataset~\cite{DeGiovanni2021}.

\subsection{\capitalisesubsection{Test benches on the heterogeneous platform}\label{sec:ad-testbenches}}
The three designs are mapped on the PULP-based Mr.Wolf platform to estimate their overall energy consumption and perform the energy-accuracy analysis. In all the test benches, the preprocessing always runs on the \gls{fc}. The first two test benches consist of 1) \algorithmName{} running on the \gls{fc} with the \gls{cl} power-gated, and 2) \adaptiveAlgName{} always running on the \gls{cl}. The third test bench consists of the fully adaptive process, including the error detection, with \algorithmName{} running on the \gls{fc} and  \adaptiveAlgName{} running on \gls{cl} when \algorithmName{} fails. Each of the test benches is applied to the $99$ segments described in Section~\ref{subsec:ad-database}. 

To measure the execution time of the three configurations, we used the open PULP platform \cite{PULPSDK}. PULP provides an SDK to run RTL simulations, using Modelsim, in order to obtain a \mbox{cycle-accurate} profiling. To estimate the energy consumption of our proposed system, we use the power numbers reported for a chip based on the PULP architecture implemented in TSMC \SI{40}{\nano\meter} LP~CMOS technology, namely, Mr.Wolf~\cite{Pullini2019}. We consider the lowest energy point of the platform, at \SI{0.8}{\volt}. The platform requires \SI{3.6}{\micro\watt}~\cite{flamand_2018_gap8} when power-gated\footnote{As reported for GAP-8~\cite{flamand_2018_gap8}, which is an industrial version of PULP with state-of-the-art deep sleep optimizations not yet included in Mr.Wolf, its academic counterpart.} and \SI{12.6}{\micro\watt} with full L2 retention. To implement better memory management of the activated banks as done in~\cite{DeGiovanni2020_TCAD}, we reduce the L2 to \SI{128}{\kibi\byte}, with a resolution of \SI{16}{\kibi\byte} per memory bank, since the application does not need more memory. 
When the \gls{soc} architecture of PULP is active, it consumes \SI{0.98}{\milli\watt} with its main processor clock-gated, and \SI{6.66}{\milli\watt} while operating at \SI{170}{\mega\hertz}.
Once the \gls{cl} is activated, it consumes \SI{0.61}{\milli\watt} with all the cores clock-gated and \SI{18.87}{\milli\watt} with the eight~cores running at \SI{110}{\mega\hertz}. 

The three designs are compared first in terms of accuracy, then energy consumption of their mapping on the PULP platform and then in their energy-accuracy trade-off for all the subjects in the dataset and as a summary for worst, average and best cases. 

\section{\capitalisesection{Experimental results}\label{sec:ad-results}}
\begin{figure*}[tp]
	\centering
	\subfloat[Worst case]{\includegraphics[width=0.3\linewidth]{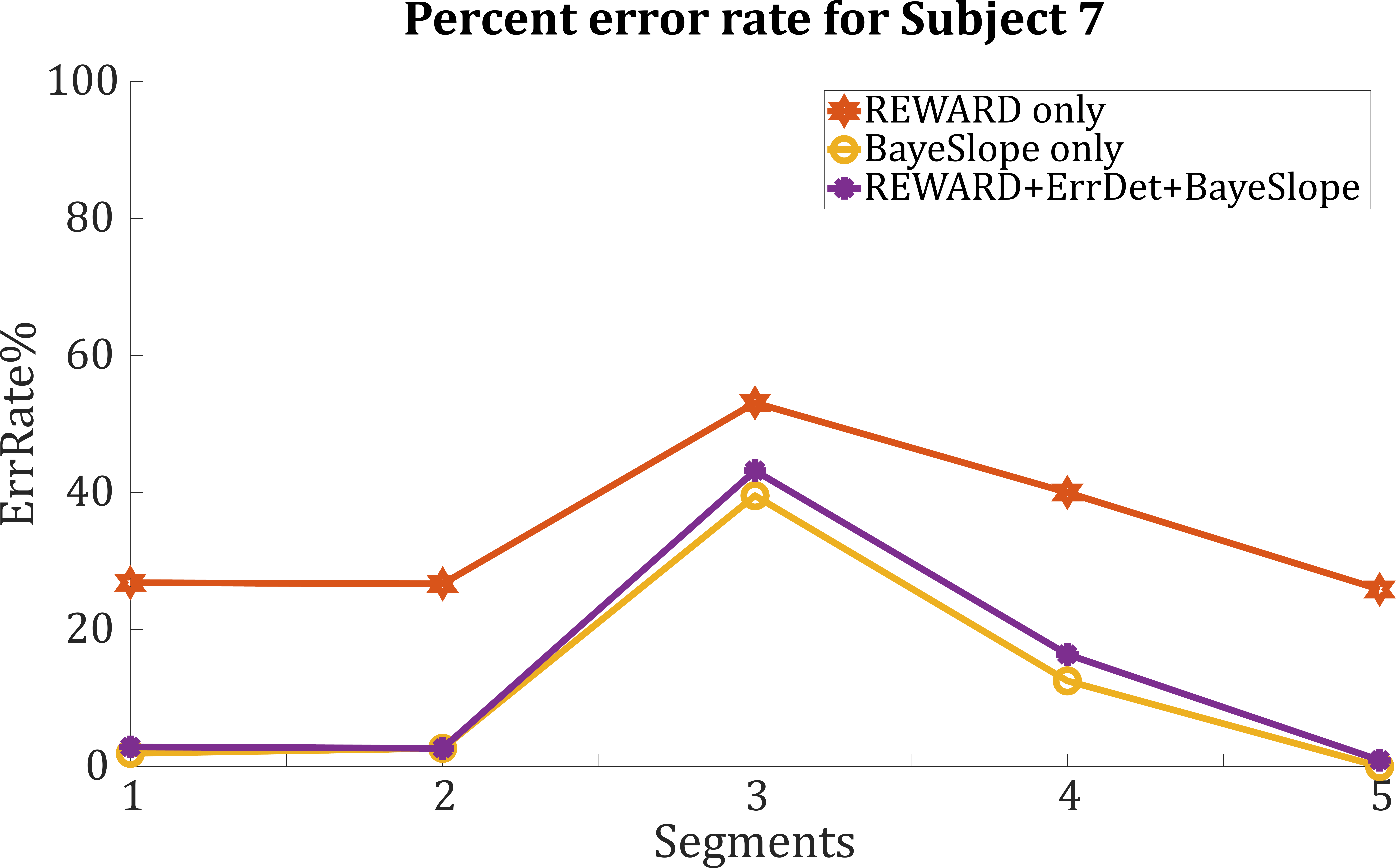} \label{fig:ad-acc1}}
	\qquad
	\subfloat[Average case]{\includegraphics[width=0.3\linewidth]{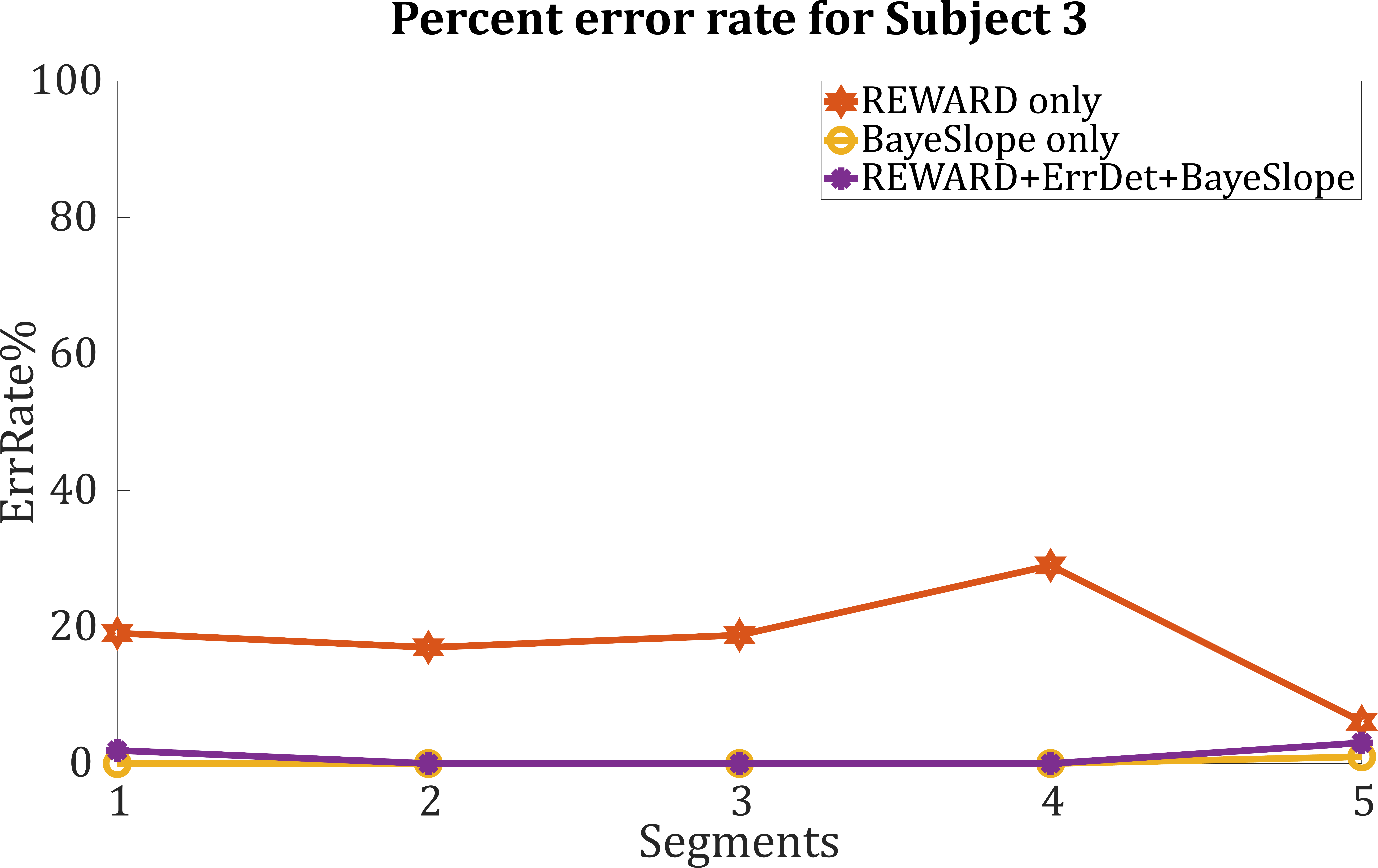}\label{fig:ad-acc2}}
	\qquad
	\subfloat[Best case]{\includegraphics[width=0.3\linewidth]{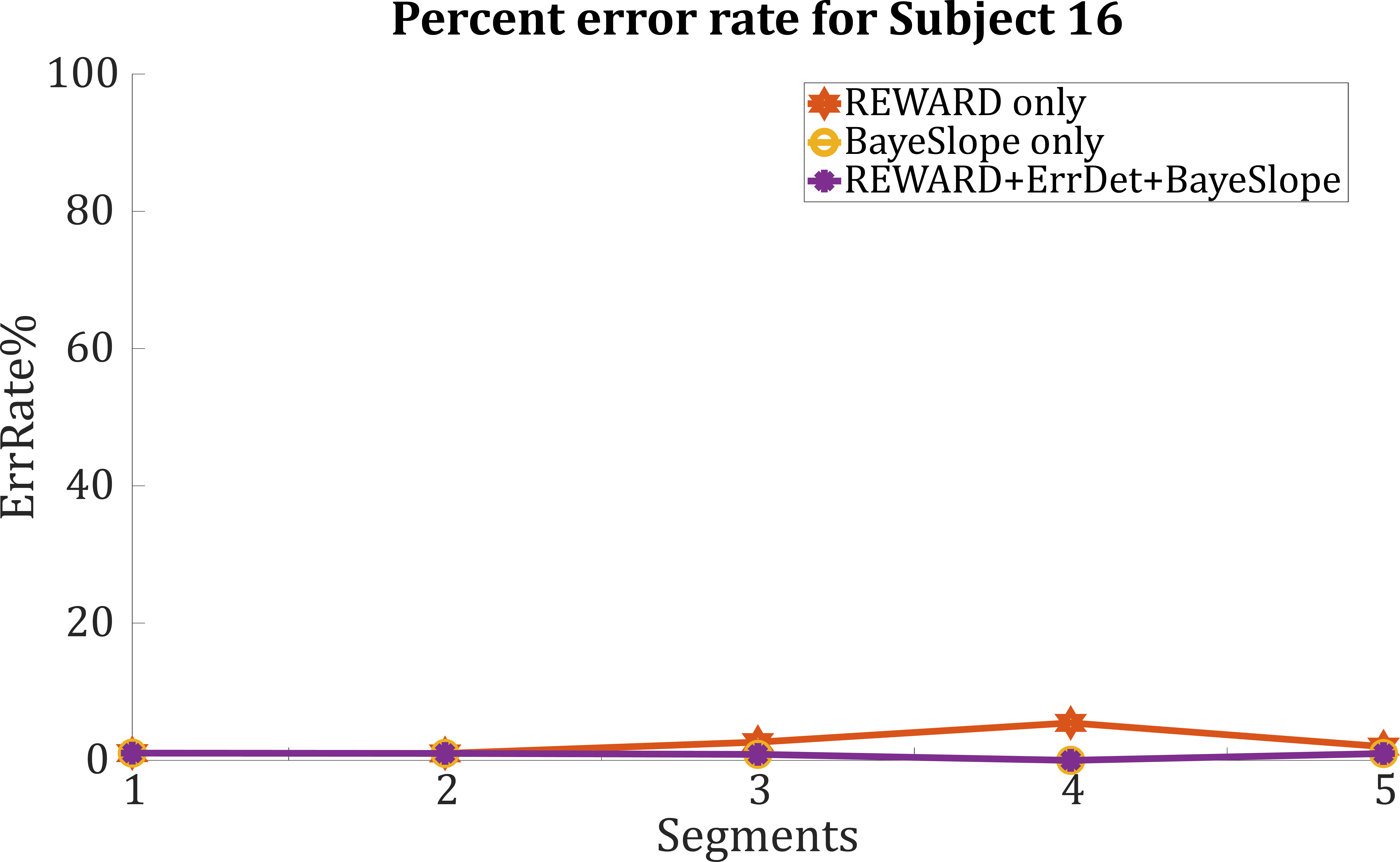}\label{fig:ad-acc3}}		
	\caption[Percent error rate of \algorithmName{}, \adaptiveAlgName{} and the online adaptive design]{Percent error rate with respect to the manual annotations of the three designs described in Section~\ref{subsec:ad-database}, for three worst, average, and best case subjects along five increasing exercise intensities. Segments 1 and 2 represent the exercise before and after VT2, segments 3 and 4 before and during exhaustion (VO2max), and segment 5 is the recovery phase right after exhaustion}
	\label{fig:ad-accuracy}
\end{figure*}

\subsection{\capitalisesubsection{Accuracy analysis of the test benches}\label{subsec:ad-accuracy}}
In Fig.~\ref{fig:ad-accuracy}, we report the percent of the error rate (ErrRate\%) in the peak detection of the three designs, described in Section~\ref{subsec:ad-database}, and its evolution through the type of segments for three example subjects. These examples illustrate three cases within the worst, best, and average groups in terms of accuracy of the new algorithm, \adaptiveAlgName{}, and the fully adaptive design (\algorithmName{}+Error detection (ErrDet)+\adaptiveAlgName{}) compared to \algorithmName{}. ErrRate\% is computed as (1-$F_1$)*100, where $F_1$ score is a measure of the peak detection performance defined as
\begin{equation}
 F_1 = \frac{TP}{TP + \frac{1}{2}\cdot(FP+FN)}
\end{equation}
TP is the set of the correctly detected peaks that match the manual annotations. FP represents all the misdetected peaks by the algorithm. FN is the set of all the peaks in the algorithm that do not match any manual annotation. The different segments shown in Fig.~\ref{fig:ad-accuracy} represent increasing exercise intensities till the recovery after exhaustion (segment~5), as described in Section~\ref{sec:ad-expsetup}. In Fig.~\ref{fig:ad-acc1}, Subject~7 has one of the worst error rates for the new algorithm, and the reason is that segment~3 is quite noisy. 
\begin{figure}[tp]
	\centering
	\includegraphics[width=1\linewidth]{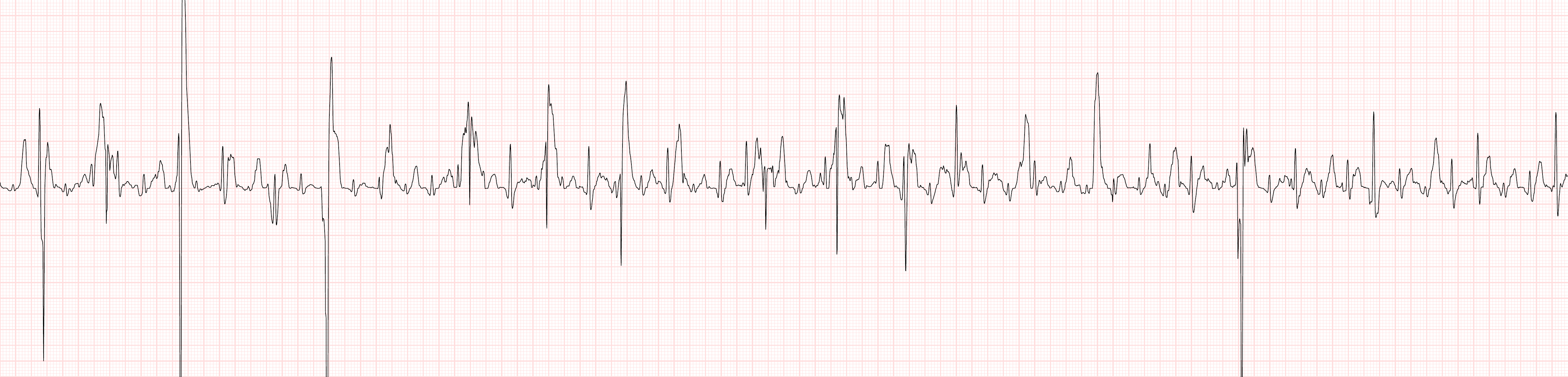}
	\caption[ECG extracted from worst case subject acquired]{\gls{ecg} segment~3 (i.e., before VT2) for Subject~7. The amplitude of the peaks is highly variable due to the changes in the exercise intensities.  The signal is shown on a standard \gls{ecg} sheet containing small squares of \SI{1}{\milli\meter}$\cdot$\SI{1}{\milli\meter} corresponding to \SI{40}{\milli\second} (horizontal) and \SI{0.1}{\milli\volt} (vertical)~\cite{ThalerECG2018}. They also include big squares of \SI{5}{\milli\meter}$\cdot$\SI{5}{\milli\meter} and correspond to \SI{200}{\milli\second}$\cdot$\SI{0.5}{\milli\volt}.}
	\label{fig:ecg_sub7_exc3}
\end{figure}
The quality of the segment is shown in Fig.~\ref{fig:ecg_sub7_exc3}, where the amplitude of the \gls{ecg} has a high variability due to changes caused by the exercise intensities near exhaustion (segment~3 is before VO$_2$max). However, \adaptiveAlgName{} and its adaptive design, with an $F_1$ score at approximately \SI{60.5}{\percent} and \SI{56.8}{\percent}, respectively, gains within \SI{13.5}{\percent} and \SI{9.9}{\percent} in performance, compared to \algorithmName{}. In Fig.~\ref{fig:ad-acc2}, Subject~3 represents an average case where \algorithmName{} has a lower error rate compared to the worst case (Subject~7), though still significant. In fact, the adaptive design performs significantly better, with an error rate up to \SI{3}{\percent}, slightly worse than \adaptiveAlgName{}. In Fig.~\ref{fig:ad-acc3}, Subject~16 is one of the best cases where \algorithmName{} fails only during more intense exercise (at exhaustion), with an error rate up to \SI{5.5}{\percent}, while \adaptiveAlgName{} has an error rate of only \SI{1}{\percent}. 

Considering the five exercise intensities, a relevant summary of the algorithms' performance is depicted in Table~\ref{tbl:accuracy-adaptive}. Here, we report the $F_1$ score, sensitivity, and \gls{ppv} of the three test benches for each of the five types of segment computed across the subjects, as well as the mean and standard deviation of the time difference between each test bench output and the manual annotations. Moreover, we report the values of the standard GQRS detection algorithm~\cite{WFDB_Moody2021}. \adaptiveAlgName{} is the most accurate of the three designs over all the performance parameters. In terms of $F_1$ score, it is comparable to the GQRS detection during lower intensity exercises (up to \SI{0.3}{\percent} before and after VT2, and \SI{0.7}{\percent} during the recovery after exhaustion where the \gls{ecg} stabilizes), while it performs better during intense physical exercise (up to \SI{8.4}{\percent} in the rest of the segments), due to the lower sensitivity of GQRS. In fact, by adapting to all the changes in the \gls{ecg} during intense physical exercise, it reaches an $F_1$ score up to \SI{99.3}{\percent} with very low variability through the segments. On the contrary, the $F_1$ score and the sensitivity of \algorithmName{} during more intense exercise (before and at VO$_2$max), where sudden changes in \gls{ecg} occur, are significantly lower than the acceptable medical standard, compared to less intense exercise. However, combining both methods in an adaptive design is as accurate as \adaptiveAlgName{} (up to \SI{1.7}{\percent} of difference in $F_1$ score). 

\begin{table*}[tp]
	\caption[Performance scores of the proposed \algorithmName{}, \adaptiveAlgName{} and the online adaptive design]{$F_1$ score, \gls{ppv}, sensitivity (\%) for the three test benches and the five exercise intensities computed across the subjects}\label{tbl:accuracy-adaptive}
	\centering
	\begin{tabular}{ll|rrrrr|r}
		\toprule
		& & \multicolumn{1}{c|}{Before VT2}             & \multicolumn{1}{c|}{After VT2}              & \multicolumn{1}{c|}{Before VO$_2$max}              & \multicolumn{1}{c|}{VO$_2$max}              & \multicolumn{1}{c|}{Recovery} & \multicolumn{1}{c}{\textbf{Total}} \\ \cmidrule{1-8} 
		\multicolumn{1}{l|}{\multirow{5}{*}{$F_1$ (\%)}} & GQRS &
		\multicolumn{1}{c|}{99.3}  & \multicolumn{1}{c|}{99.1}  & \multicolumn{1}{c|}{89.5} &  \multicolumn{1}{c|}{91.3} & \multicolumn{1}{c|}{98.6} & \multicolumn{1}{c}{95.4} \\ \cmidrule{2-8} 	
		\multicolumn{1}{l|}{} & \algorithmName\space (RW)& \multicolumn{1}{c|}{92.1}  & \multicolumn{1}{c|}{90.9}  & \multicolumn{1}{c|}{78.7} &  \multicolumn{1}{c|}{80.2} & \multicolumn{1}{c|}{92.5} & \multicolumn{1}{c}{86.7} \\ 
		\multicolumn{1}{l|}{} & \adaptiveAlgName\space (BS) &  \multicolumn{1}{c|}{\textbf{99.0}} & \multicolumn{1}{c|}{\textbf{99.1}} &  \multicolumn{1}{c|}{\textbf{97.9}} & \multicolumn{1}{c|}{\textbf{98.8}}  &  \multicolumn{1}{c|}{\textbf{99.3}} & \multicolumn{1}{c}{\textbf{98.8}} \\ 
		\multicolumn{1}{l|}{} & RW + ErrDet + BS &  \multicolumn{1}{c|}{98.9} & \multicolumn{1}{c|}{99.0}  & \multicolumn{1}{c|}{96.2}  & \multicolumn{1}{c|}{97.1}  &  \multicolumn{1}{c|}{98.5} & \multicolumn{1}{c}{97.9} \\ \cmidrule{1-8} 
		\multicolumn{1}{l|}{\multirow{5}{*}{\gls{ppv} (\%)}}  & GQRS & 
		\multicolumn{1}{c|}{99.6}  & \multicolumn{1}{c|}{99.5}  & \multicolumn{1}{c|}{99.4} &  \multicolumn{1}{c|}{99.7} & \multicolumn{1}{c|}{100.0} & \multicolumn{1}{c}{99.6} \\ \cmidrule{2-8} 	
		\multicolumn{1}{l|}{} & \algorithmName\space (RW)& \multicolumn{1}{c|}{98.2}  & \multicolumn{1}{c|}{98.2}  & \multicolumn{1}{c|}{97.1} &  \multicolumn{1}{c|}{96.3} & \multicolumn{1}{c|}{98.1} & \multicolumn{1}{c}{97.6} \\ 
		\multicolumn{1}{l|}{} & \adaptiveAlgName\space (BS) &  \multicolumn{1}{c|}{\textbf{98.6}} & \multicolumn{1}{c|}{\textbf{98.6}} &  \multicolumn{1}{c|}{\textbf{98.9}} & \multicolumn{1}{c|}{\textbf{98.6}}  &  \multicolumn{1}{c|}{\textbf{98.6}} & \multicolumn{1}{c}{\textbf{98.7}} \\ 
		\multicolumn{1}{l|}{} & RW + ErrDet + BS &  \multicolumn{1}{c|}{98.3} & \multicolumn{1}{c|}{98.4}  & \multicolumn{1}{c|}{97.3}  & \multicolumn{1}{c|}{96.2}  &  \multicolumn{1}{c|}{97.5} & \multicolumn{1}{c}{97.5} \\ \cmidrule{1-8} 
		\multicolumn{1}{l|}{\multirow{5}{*}{Sensitivity (\%)}}  & GQRS &
		\multicolumn{1}{c|}{99.0}  & \multicolumn{1}{c|}{98.7}  & \multicolumn{1}{c|}{81.4} &  \multicolumn{1}{c|}{84.2} & \multicolumn{1}{c|}{97.2} & \multicolumn{1}{c}{91.6} \\ \cmidrule{2-8} 	
		\multicolumn{1}{l|}{} & \algorithmName\space (RW)& \multicolumn{1}{c|}{86.8}  & \multicolumn{1}{c|}{84.5}  & \multicolumn{1}{c|}{66.1} &  \multicolumn{1}{c|}{68.7} & \multicolumn{1}{c|}{87.4} & \multicolumn{1}{c}{78.0} \\
		\multicolumn{1}{l|}{} & \adaptiveAlgName\space (BS) &  \multicolumn{1}{c|}{99.3} & \multicolumn{1}{c|}{99.5} &  \multicolumn{1}{c|}{\textbf{96.9}} & \multicolumn{1}{c|}{\textbf{98.9}}  &  \multicolumn{1}{c|}{\textbf{100.0}} & \multicolumn{1}{c}{\textbf{98.9}}\\ 
		\multicolumn{1}{l|}{} & RW + ErrDet + BS &  \multicolumn{1}{c|}{\textbf{99.4}} & \multicolumn{1}{c|}{\textbf{99.6}}  & \multicolumn{1}{c|}{95.2}  & \multicolumn{1}{c|}{98.0}  &  \multicolumn{1}{c|}{99.6} & \multicolumn{1}{c}{98.3} \\ 
		\cmidrule{1-8} 	
		\multicolumn{1}{l|}{\multirow{5}{*}{\begin{tabular}[c]{@{}l@{}}Time (ms) \\ from manual \\ annotation \end{tabular}}}  & GQRS &
		\multicolumn{1}{c|}{0.2$\pm$3.7}  & \multicolumn{1}{c|}{0.4$\pm$4.7}  & \multicolumn{1}{c|}{5.5$\pm$23.6} &  \multicolumn{1}{c|}{8.1$\pm$29.2} & \multicolumn{1}{c|}{3.2$\pm$20.4} & \multicolumn{1}{c}{3.4$\pm$19.4} \\ \cmidrule{2-8} 	
		\multicolumn{1}{l|}{} & \algorithmName\space (RW)& \multicolumn{1}{c|}{0.6$\pm$8.4}  & \multicolumn{1}{c|}{1.1$\pm$11.2}  & \multicolumn{1}{c|}{10.4$\pm$35.7} &  \multicolumn{1}{c|}{20.6$\pm$48.6} & \multicolumn{1}{c|}{9.2$\pm$34.8}  & \multicolumn{1}{c}{8.1$\pm$32.0} \\ 
		\multicolumn{1}{l|}{} & \adaptiveAlgName\space (BS) &  \multicolumn{1}{c|}{\textbf{0.5$\pm$6.5}} & \multicolumn{1}{c|}{\textbf{0.3$\pm$4.6}} &  \multicolumn{1}{c|}{4.9$\pm$24.0} & \multicolumn{1}{c|}{\textbf{6.8$\pm$29.1}}  &  \multicolumn{1}{c|}{\textbf{1.0$\pm$10.3}} & \multicolumn{1}{c}{2.9$\pm$18.6} \\ 
		\multicolumn{1}{l|}{} & RW + ErrDet + BS &  \multicolumn{1}{c|}{\textbf{0.5$\pm$6.5}} & \multicolumn{1}{c|}{\textbf{0.3$\pm$4.6}}  & \multicolumn{1}{c|}{\textbf{4.3$\pm$22.3}}  & \multicolumn{1}{c|}{7.3$\pm$30.1}  &  \multicolumn{1}{c|}{1.1$\pm$11.3} & \multicolumn{1}{c}{\textbf{2.8$\pm$18.5}} \\ \bottomrule
	\end{tabular}
\end{table*}

\begin{figure}[tp]
	\centering
	\includegraphics[width=0.9\linewidth]{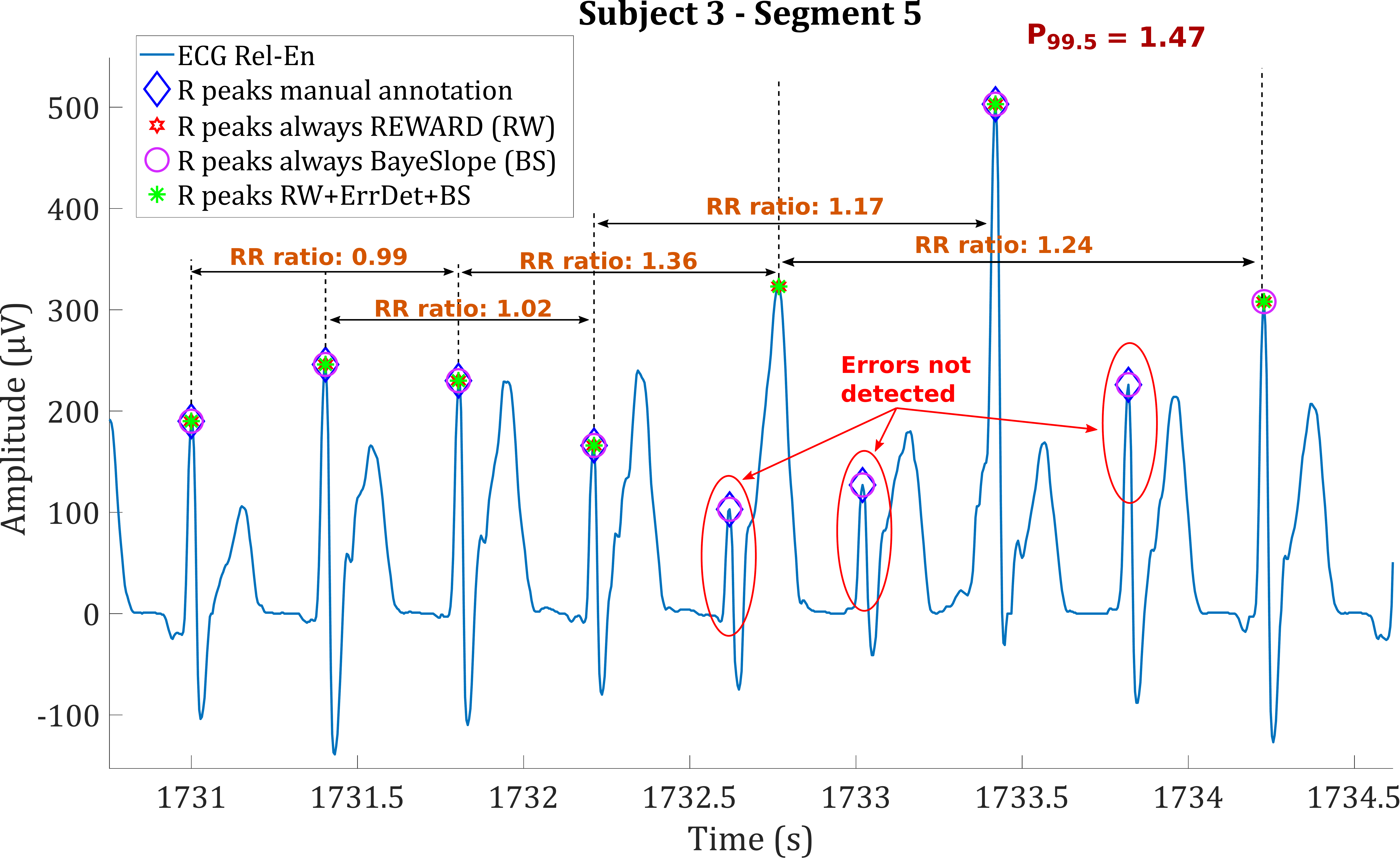}
	\caption[Example of error detection failing]{\gls{ecg} segment~5 (i.e., recovery after exhaustion) for Subject~3 with the R peaks from the three designs. The RR ratio where the small R peaks are not detected by \algorithmName{} is close to $P_{99.5}$ but not enough to trigger an error.}
	\label{fig:sub3_exc5_errdetwrong}
\end{figure}

Rarely, the adaptive design could perform better (less than \SI{1}{\percent} difference in score) as it is shown in the sensitivity values. This is due to the initialization process of \adaptiveAlgName{}, which requires the signal to be stable as it does not use any prior information within this initial stage. Therefore, it happens rarely that the signal is more stable later in the segment where \adaptiveAlgName{} is triggered and will be initialized, compared to the initialization at the beginning of the segment (when always running \adaptiveAlgName{}). This can also cause a delay in the adaptation and very few peaks missed and result instead in a slightly worse accuracy. Another reason for a lower performance in the adaptive design compared to always running \adaptiveAlgName{}, specifically for more intense exercise (before and after VO$_2$max) and during recovery, as shown in Table~\ref{tbl:accuracy-adaptive}, is due to an issue in the error detection. In fact, the RR ratio distribution used to compute the tail thresholds is performed on the full dataset and accounting for different exercise intensities. Within more intense exercises, as the RR intervals get smaller, it can happen that even if \algorithmName{} misses one peak, the RR ratio is still within the distribution. This is shown in Fig.~\ref{fig:sub3_exc5_errdetwrong}, where the RR ratio computed on the small peaks not detected by \algorithmName{} is close to the $P_{99.5}$ of the distribution but not enough to trigger an error. This results in a lower accuracy for the adaptive design.
One way to fix this problem is to compute different distributions for different exercise intensities. In the case of this dataset, it could be five distributions or two groups of low and high intensities. Another way is to adapt the distribution online by detecting the intensity type and choose the correct tail thresholds. Finally, with the more advanced capabilities of modern heterogeneous platforms, the distribution can be computed directly on the signal acquired through a small training process on \adaptiveAlgName{} and then adapting the tail thresholds. 

In conclusion, the accuracy results show that always running \adaptiveAlgName{} is the most accurate and robust of the three designs. At the same time, \algorithmName{}'s performance highly varies with the intensity of the exercise. However, \adaptiveAlgName{} is approximately \SI{100}{\times} more complex than the R peak detection step of \algorithmName{}. Therefore, we propose the adaptive design that combines both algorithms and has a similar accuracy compared to \adaptiveAlgName{}. In the next section, we will show the advantages in terms of energy consumption of the adaptive design on the PULP platform.  

\subsection{\capitalisesubsection{Energy consumption of test benches in PULP}}
\begin{figure*}[tp]
	\centering
	\subfloat[Worst case]{\includegraphics[width=0.3\linewidth]{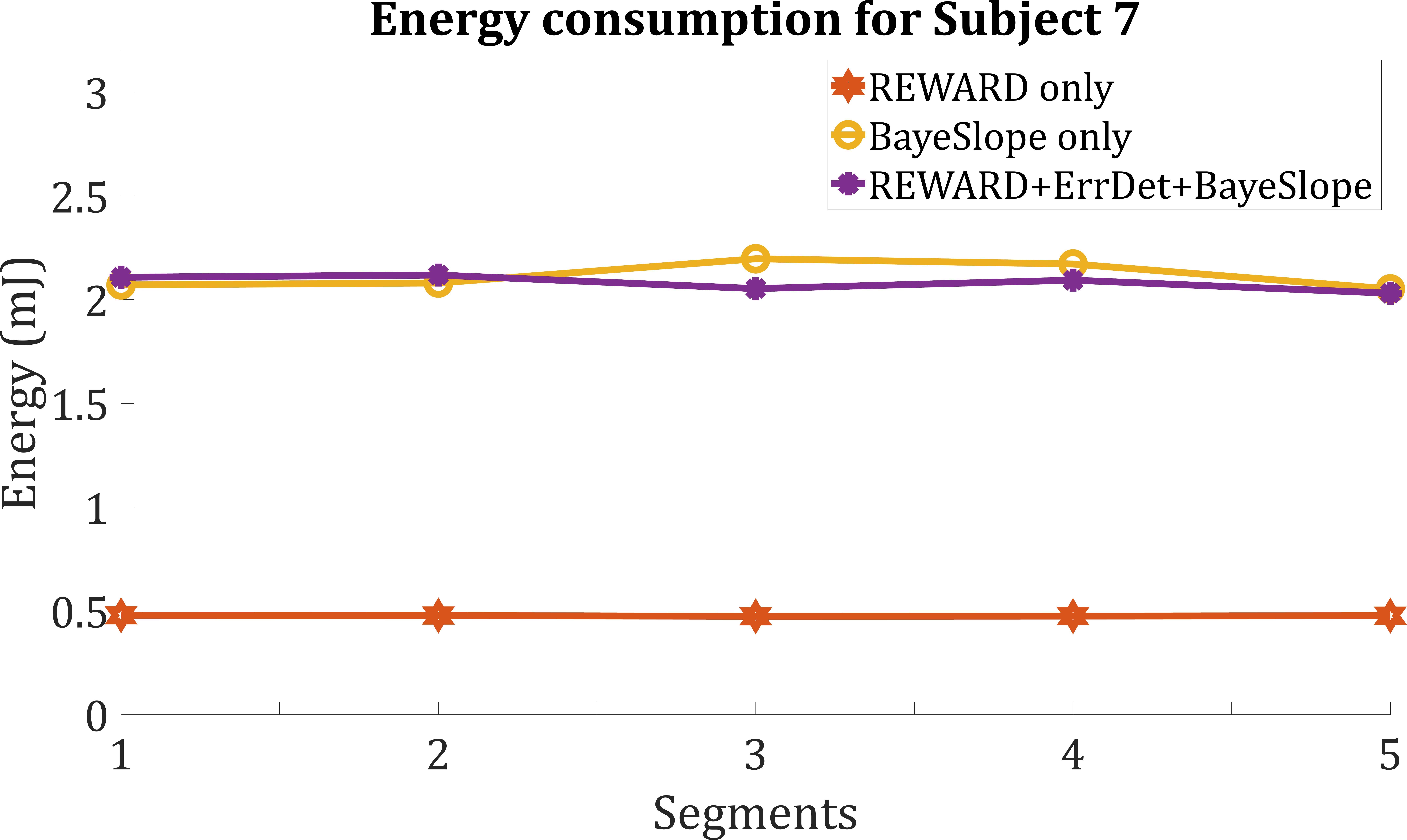}\label{fig:ad-energy1}}
	\qquad
	\subfloat[Average case]{\includegraphics[width=0.3\linewidth]{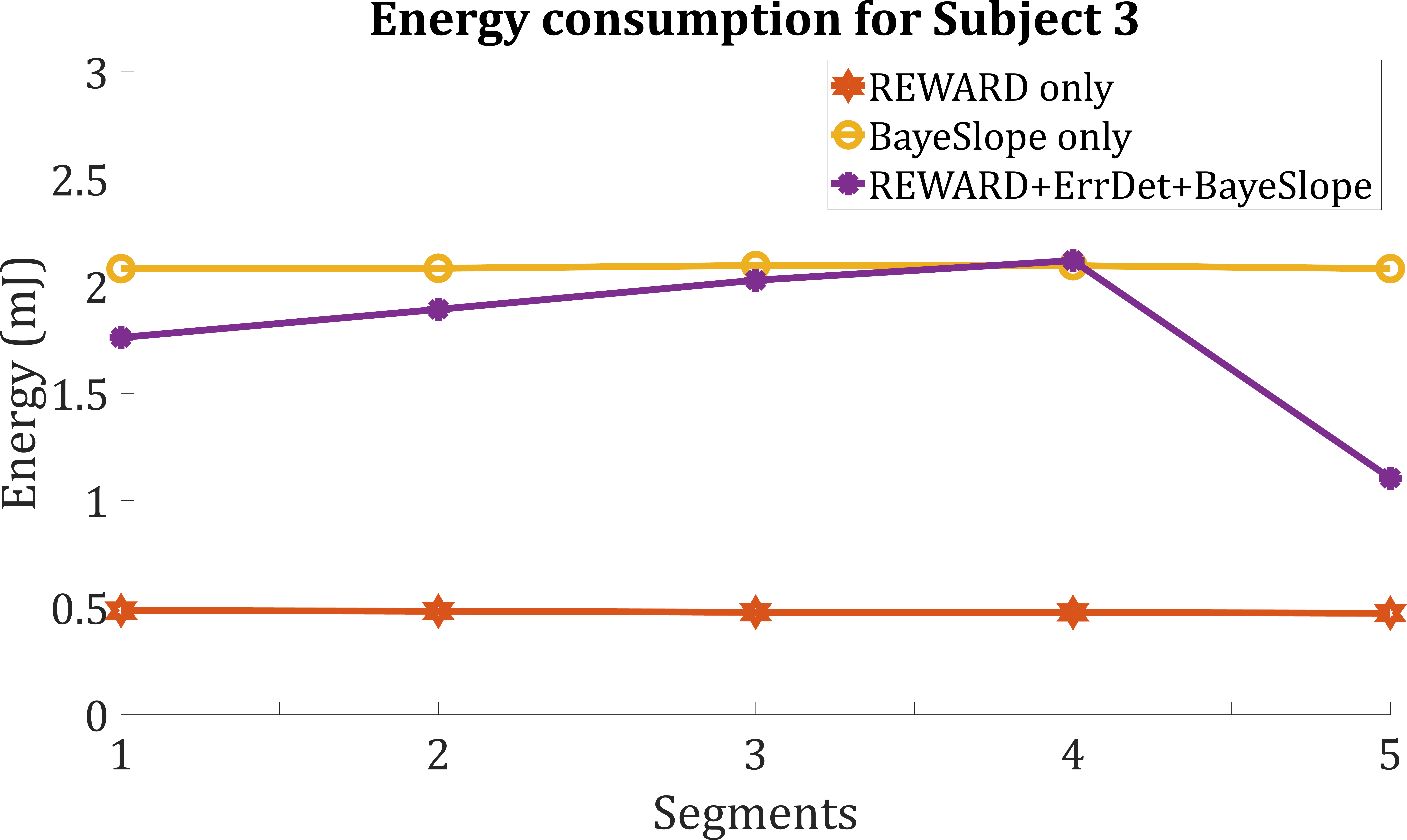}\label{fig:ad-energy2}}
	\qquad
	\subfloat[Best case]{\includegraphics[width=0.3\linewidth]{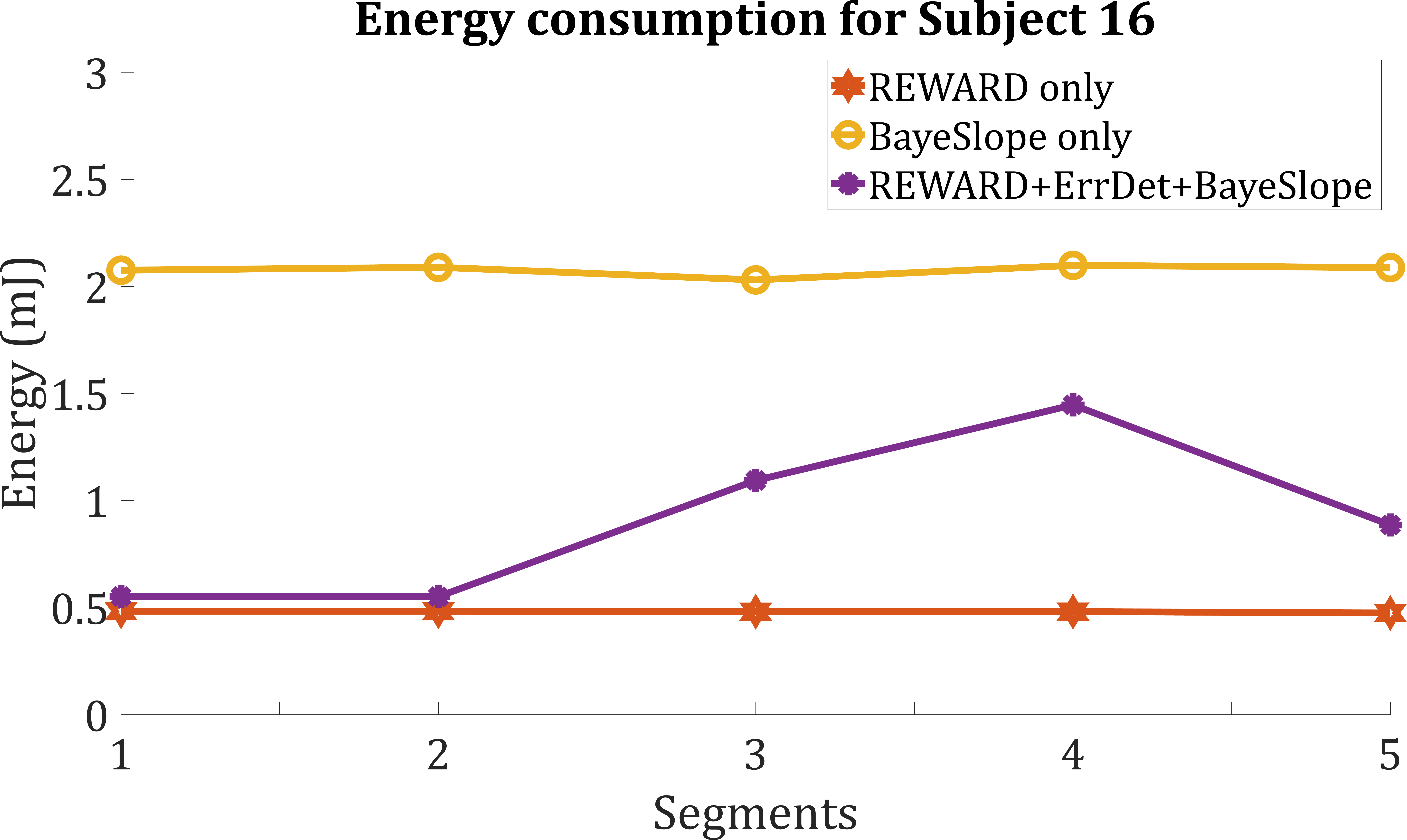}\label{fig:ad-energy3}}		
	\caption[Energy consumption of \algorithmName{}, \adaptiveAlgName{} and the online adaptive design]{Energy consumption of the three test benches described in Section~\ref{sec:ad-testbenches} for three worst, average, and best case subjects along five increasing exercise intensities. Segments 1 and 2 represent the exercise before and after VT2, segments 3 and 4 before and during exhaustion (VO2max), and segment 5 is the recovery right after exhaustion}
	\label{fig:ad-energy}
\end{figure*}
Figure~\ref{fig:ad-energy} shows the energy consumption of the platform for the three subjects described in Section~\ref{subsec:ad-accuracy}. In Subject~7 (Fig.~\ref{fig:ad-energy1}), the worst case scenario, the fully adaptive design consumes the same amount of energy in almost all the windows. In segment~3, the adaptive design achieves \SI{6.5}{\percent} of energy savings compared to always running \adaptiveAlgName{}, with a \SI{3.7}{\percent} difference in $F_1$ score. However, the overall accuracy is far from the required medical standard, even the state-of-the-art GQRS algorithm, where it reaches only \SI{75}{\percent}. In Subject~3 (Fig.~\ref{fig:ad-energy2}), for all the exercise intensities except segment~4, during exhaustion, the fully adaptive wearable design we propose has energy savings up to \SI{48}{\percent} compared to \adaptiveAlgName{} with a loss in accuracy of only up to \SI{2}{\percent} (c.f.~Fig.~\ref{fig:ad-acc2}). For segment~3, even if the energy savings are one of the lowest at approximately \SI{3.3}{\percent}, the fully adaptive design is as accurate as \adaptiveAlgName{} and \SI{18.8}{\percent} more accurate compared to \algorithmName{}. Therefore, on average cases such as Subject~3, in most exercise intensities, choosing the fully adaptive design can improve the energy-accuracy trade-off. Subject~16, representing one of the best case scenarios in Fig.~\ref{fig:ad-energy3}, highlights the adaptivity of the full design and its error detection through the segments, starting with a minimum energy consumption, since only \algorithmName{} is running, and maximum attainable accuracy. Then, when the exercise intensity increases, \algorithmName{} fails more frequently, and \adaptiveAlgName{} takes over the R peak detection. Finally, during recovery, when the \gls{ecg} stabilizes and \algorithmName{} fails less compared to exhaustion, the energy consumption drops to a lower level. Our fully adaptive design maintains a high level of accuracy (approximately \SI{99}{\percent}), while limiting the energy consumption compared to  executing \adaptiveAlgName{} for the full segment, with energy savings from \SI{31.8}{\percent} up to \SI{58.6}{\percent}. 
\begin{figure}[tp]
	\centering
	\includegraphics[width=1\linewidth]{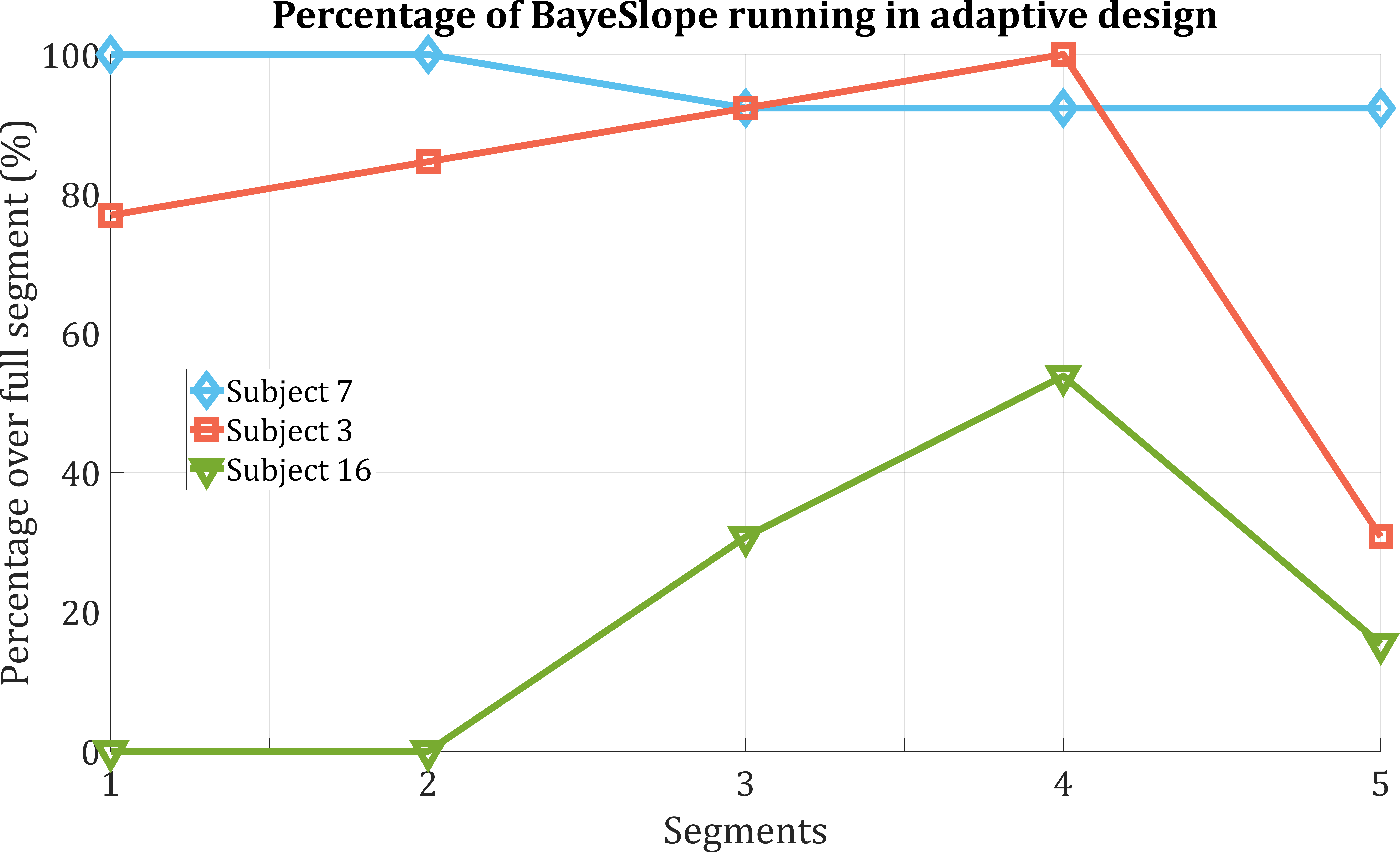}	
	\caption[Percentage of windows running \adaptiveAlgName{} in the online adaptive design]{Percentage of windows over the full segment where \adaptiveAlgName{} is triggered during the adaptive design for three worst, average, and best case scenarios. Comparing these trends with the ones shown in Fig.~\ref{fig:ad-energy}, it is evident that the adaptive design reduces energy consumption by reducing the number of times \adaptiveAlgName{} runs on the \gls{cl}} 
	\label{fig:ad-percBStrigger}
\end{figure}

Fig.~\ref{fig:ad-percBStrigger} shows how many times \adaptiveAlgName{} runs in the adaptive design in terms of percentage of windows over the full segment for the three cases analyzed. Considering the windows where an error occurs and triggers \adaptiveAlgName{}, the previous window also counts as triggered since \adaptiveAlgName{} needs an additional window for the initialization process (c.f.,~Section~\ref{ad-subsec:adapt_platform}). As expected, the trend is similar to the energy reduction compared to always running \adaptiveAlgName{} shown in Fig.~\ref{fig:ad-energy}. However, for Subject~7, starting from segment~3, the trend is slightly different than the one in Fig.~\ref{fig:ad-energy1}. In fact, segment~3 has a \SI{6.5}{\percent} reduction in energy of the adaptive design compared to \adaptiveAlgName{}, while it is less than \SI{1}{\percent} in segment~4. This occurs because \adaptiveAlgName{} includes an overlapping for every window to avoid missing peaks at the border between two windows. This does not occur in the adaptive design (segment~3) when the algorithm runs once on two windows. Thus, it avoids this overhead with a small advantage in energy consumption. 
% \begin{figure}[tp]
% 	\centering
% 	\subfloat[Excerpt 3]{\includegraphics[width=0.9\linewidth]{figures/error_freq_sub7_exc3.pdf}\label{fig:ad-err_exc3}}
% 	\qquad
% 	\subfloat[Excerpt 4]{\includegraphics[width=0.9\linewidth]{figures/error_freq_sub7_exc4.pdf}\label{fig:ad-err_exc4}}	
% 	\caption[Error occurrence in two ECG excerpts from the same subject]{Error occurrence in two excerpts of Subject 7 with the same percentage of \adaptiveAlgName{} triggers shown in Fig.~\ref{fig:ad-percBStrigger}. Excerpt 3 has an alternating pattern, while in excerpt 4 the error occurs in consecutive windows, which explains the slightly bigger energy reduction in excerpt 3 compared to expert 4}
% 	\label{fig:ad-errpattern}
% \end{figure}
% To explain this result, we drew the error detection pattern in excerpt~3 and 4 for Subject~7 shown in Fig.~\ref{fig:ad-errpattern}, where 1 indicates that an error occurred. In excerpt~3, the error does not happen on every window but alternating, although \adaptiveAlgName{} runs for two windows whenever the error in the previous window is 0. On the contrary, in excerpt~4 the error is triggered in consecutive windows and \adaptiveAlgName{} runs as well on every window. When \adaptiveAlgName runs on every window an overlapping occurs to avoid missing peaks at the border between two windows. This does not occur in the adaptive design (excerpt~3) when the algorithm runs once on two windows, avoiding this overhead with a small advantage in energy consumption. 
The large differences between the three subjects show how the proposed design can adapt to the subject and different exercise intensities to reduce energy consumption instead of constantly falling in the worst case scenario. This personalized and adaptive reduction in energy consumption can lead to a longer battery lifetime for \glspl{wsn} and better usability.  

\begin{table}[tp]
	\caption[Energy consumption of the proposed \algorithmName{}, \adaptiveAlgName{} and the online adaptive design]{Energy consumption in mJ for the three test benches and the five exercise intensities computed across the subjects}\label{tbl:energy-adaptive}
	\centering
	\begin{tabular}{ll|c|c|c}
		\toprule
		&                & \begin{tabular}[c]{@{}c@{}} \algorithmName{}\\ (RW)  \end{tabular}                & \begin{tabular}[c]{@{}c@{}} \adaptiveAlgName{}\\ (BS)  \end{tabular}                   & \begin{tabular}[c]{@{}c@{}}RW + \\ ErrDet + \\ BS\end{tabular}      \\ \midrule
		\multicolumn{1}{l|}{\multirow{9}{*}{\begin{tabular}[l]{@{}c@{}}Energy\\(mJ)\end{tabular}}} & Before VT2     & 0.479$\pm$0.004          & 2.078$\pm$0.016          & 1.348$\pm$0.573          \\ \cmidrule{2-5}
		\multicolumn{1}{l|}{} & After VT2      & 0.479$\pm$0.003 & 2.070$\pm$0.032 & 1.469$\pm$0.556 \\ \cmidrule{2-5}
		\multicolumn{1}{l|}{} & \begin{tabular}[l]{@{}l@{}}Before \\ VO$_2$max\end{tabular}     & 0.476$\pm$0.004          & 2.071$\pm$0.037          & 1.840$\pm$0.299           \\ \cmidrule{2-5}
		\multicolumn{1}{l|}{} & VO$_2$max & 0.476$\pm$0.003 & 2.075$\pm$0.032 & 1.820$\pm$0.409 \\ \cmidrule{2-5}
		\multicolumn{1}{l|}{} & Recovery  & 0.477$\pm$0.002          & 2.080$\pm$0.020          & 1.275$\pm$0.562          \\\cmidrule{2-5}
		\multicolumn{1}{l|}{} & \textbf{Total} & \textbf{0.477$\pm$0.004} & \textbf{2.075$\pm$0.028} & \textbf{1.553$\pm$0.536} \\ \bottomrule
	\end{tabular}
\end{table}

Table~\ref{tbl:energy-adaptive} shows a summary of the average energy consumption for the five exercise intensities. As shown before in our accuracy analysis, higher exercise intensities require to run \adaptiveAlgName{} more often in the adaptive design. However, the algorithm achieves significant energy savings compared to always running \adaptiveAlgName{}. The reason is that as long as \adaptiveAlgName{} is not triggered, the adaptive design uses only the \gls{fc}. In that situation, the power of the platform corresponds to the power of the \gls{fc} and the leakage power of the \gls{cl}, which is significantly lower (approximately \SI{5}{\times}) than the power of the \gls{cl} executing \adaptiveAlgName{} on one of its cores, with the other ones clock-gated. Therefore, the energy consumption over the 25-second segment is reduced. As a result, the adaptive design achieves energy savings up to \SI{38.7}{\percent}, considering the average for the five exercise intensities. Moreover, it reaches up to \SI{74.2}{\percent} energy savings for the overall dataset analyzed, compared to the scenario where the \gls{cl} is always active and executes \adaptiveAlgName{}.

\subsection{\capitalisesubsection{Energy-accuracy trade-off on test benches}}
\begin{figure}[tp]
	\centering
	\includegraphics[width=1\linewidth]{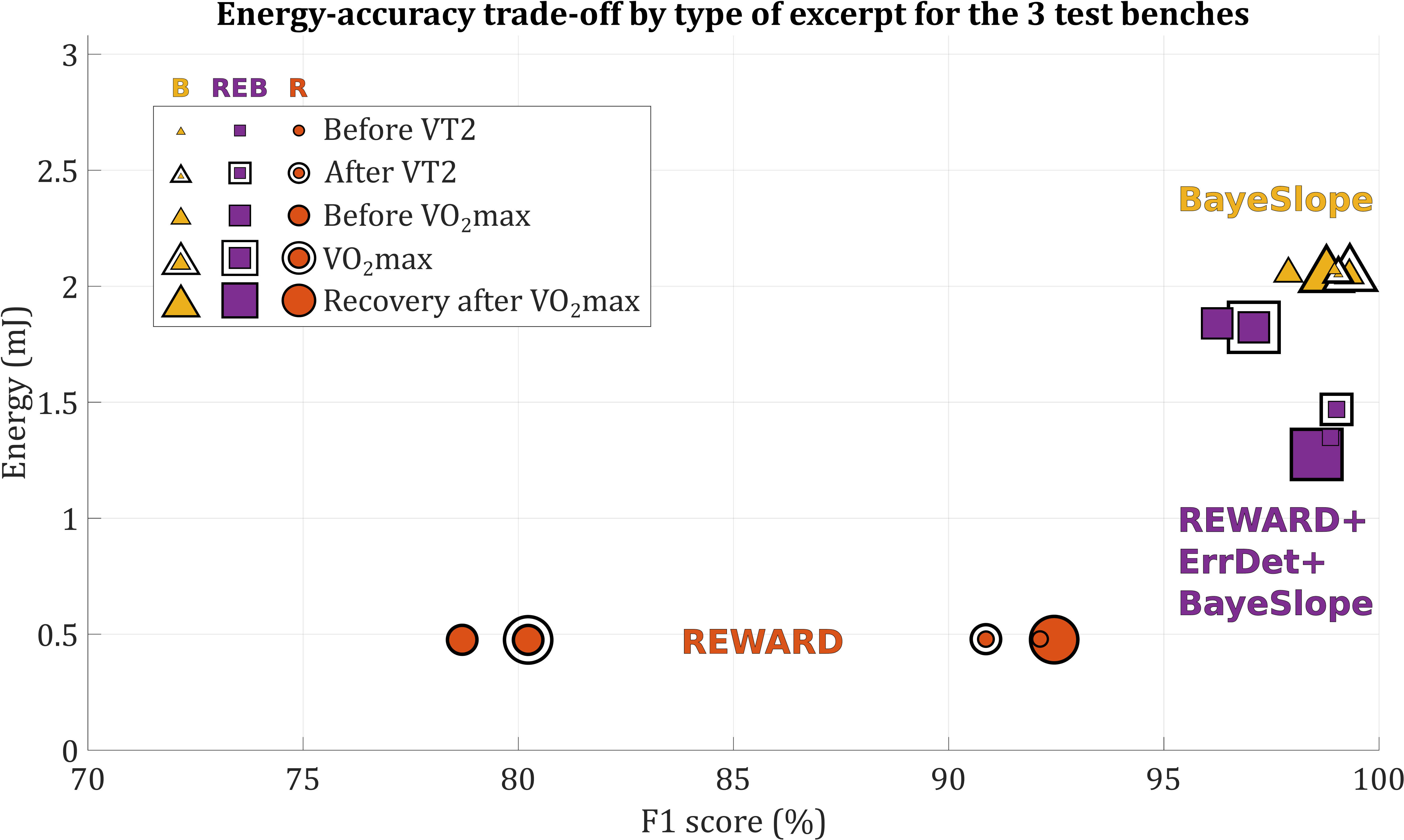}	
	\caption[Energy-accuracy analysis of \algorithmName{}, \adaptiveAlgName{} and the online adaptive design]{Energy-accuracy analysis of the three test benches and different exercise intensities}
	\label{fig:ad-energy-accuracy}
\end{figure}
Figure~\ref{fig:ad-energy-accuracy} shows the energy-accuracy comparison between the three test benches and an analysis on the different exercise intensities. We use once again the $F_1$ score as a measure of algorithm detection accuracy. For the three segments before and after VT2, and during the recovery after VO$_2$max, \algorithmName{} is accurate within the medical acceptability, and consumes the minimum energy for this application. %These values of $F_1$ score are similar to the ones we showed in the \gls{paf} application in Chapter%~\ref{ch:alg-opt}. As during exercise, \gls{paf} is characterized by sudden changes in the \gls{ecg} morphology (e.g., ectopic beats, which are usually smaller in amplitude, rhythm irregularities and missing P waves). \algorithmName{} fails to adapt to this sudden change by design, although within the medical acceptability. 
However, the fully adaptive design (in purple) is always more advantageous in terms of accuracy, with a performance increase of up to \SI{8.2}{\percent}. Moreover, it is comparable in $F_1$ score to \adaptiveAlgName{} although more energy-efficient, with energy savings up to \SI{38.7}{\percent}. 

However, when the exercise intensity increases, the number of peaks within a window increases as well. In this condition, the hysteresis thresholds of \algorithmName{} do not adapt to the high amplitude variability of the peaks within a window of analysis (\SI{1.75}{\second}), as described in Section~\ref{ad-subsec:preproc_rw_errdet} and Fig.~\ref{fig:rw_missed_peak}. In fact, before VO$_2$max the exercise intensity is about to reach its maximum, and more sudden changes in the \gls{ecg} occur, which explains the decreased accuracy of \algorithmName{}. The segment extracted during exhaustion (i.e., when reaching $VO_{2max}$) represents the highest intensity and, hence, disruption of the \gls{ecg} morphology, specifically in the amplitude of the R peak and the RR intervals (\gls{hrv} reaches its minimum). Therefore, it is the reason for a decreased performance in \algorithmName{}. On the contrary, the $F_1$ score of the fully adaptive design is only up to \SI{1.7}{\percent} lower than \adaptiveAlgName{}, which is the most accurate. The energy savings for these two segments are lower than the other three, though still significant (up to \SI{12.2}{\percent}).

Our experimental results show how the proposed \adaptiveAlgName{} algorithm is highly accurate and more robust than the lightweight \algorithmName{} when sudden changes in the \gls{ecg} morphology occur. Moreover, in these conditions, \adaptiveAlgName{} and, consequently, the adaptive design are also more robust than state-of-the-art methods such as the GQRS detector. However, if we consider the design where \adaptiveAlgName{} is mapped on a PULP-based platform and running on the \gls{cl} (with the preprocessing modules running on the \gls{fc}), the device consumes on average \SI{4.6}{\times} more than the mapping of  \algorithmName{} (and the preprocessing) in the \gls{fc}. In contrast, the adaptive design enhances the energy-accuracy trade-off, maximizing accuracy while limiting energy consumption on modern \gls{ulp} platforms. This adaptive design is not limited to applications where intense physical exercise is involved, but it can also be applied to pathologies where the \gls{ecg} morphology changes. In particular, it is applicable to pathologies beyond the wellness context, i.e., within the context of medical cardiology and aging population healthcare prevention, such as \gls{paf}~\cite{Kirchhof2016} and other types of arrhythmia. Moreover, if \adaptiveAlgName{} is parallelized in the 8-core \gls{cl}, more computing resources can be assigned to \gls{hrv} analysis and pathology detection for fully on-node processing to ensure low-rate transmission and data privacy according to the latest remote monitoring healthcare requirements. 
\section{\capitalisesection{Conclusion}\label{sec:ad-conclusion}}
In health and wellness monitoring, specifically on the cardiovascular context using wearable systems, there exist multiple pathologies and physical conditions where sudden changes in the measured biosignals occur. In particular, during intense physical exercise, sudden changes in the \gls{ecg} heart beats amplitude and rhythm cause errors in state-of-the-art standard R peak detection algorithms and, therefore, on any further analysis based on the \gls{hr}. Moreover, more accurate algorithms often require a higher amount of computing resources leading to a need for more capable wearable platforms with flexible resource management approaches. 

In this work, we have proposed a new online machine learning-based design to detect R peaks in a single-lead \gls{ecg} signal, which adapts at run time to the changes in its morphology. Furthermore, this adaptive design exploits the core heterogeneity of modern \gls{ulp} wearable platforms, which can run efficiently more complex algorithms using different types of cores. Our new online adaptive design uses a standard lightweight algorithm, \algorithmName{}, and an error detection method to measure the algorithm's accuracy. When \algorithmName{} fails, a novel algorithm called \adaptiveAlgName{}, which focuses on robustness to sudden variations in the signal properties though more complex, is triggered and runs in a more capable core. In the context of a maximal exercise test, and, in particular, during high intensity exercise, our proposed \adaptiveAlgName{} outperforms the state-of-the-art standard algorithms, such as the GQRS  detector, of up to \SI{8.4}{\percent} in $F_1$ score. Similarly, our innovative online adaptive design achieves a high $F_1 $ score, up to \SI{99.0}{\percent} across five different exercise intensities, which is comparable to always running \adaptiveAlgName{}, and up to \SI{17.5}{\percent} more accurate compared to running only \algorithmName{}. By implementing the newly proposed adaptive method in the heterogeneous PULP \gls{soc} wearable architecture, it can reach energy savings up to \SI{38.7}{\percent} compared to always running the more complex \adaptiveAlgName{}. Therefore, the newly proposed online adaptive design maximizes the accuracy while minimizing the energy consumption for an optimal energy-accuracy trade-off when used in latest \gls{soc} architectures of wearable systems.   

\section*{Acknowledgment}

The authors acknowledge the help of Dr. Fabio Montagna for the initial implementation of BayeSlope in the PULP platform, and Dr. Miguel P\'eon-Quir\'os for the help in the design of the experiments in PULP and the insightful comments on the paper. Additionally, the authors acknowledge Dr. Nicolas Bourdillon and Leandre Tschanz for the help in the preparation of the ethical protocol and the collection of the data.

\printglossary

\bibliographystyle{IEEEtran}
\bibliography{refs}

\end{document}